\documentclass[a4paper,11pt]{article}
\pdfoutput=1
\usepackage[no-natbib-sort]{jheppub}

\usepackage[utf8]{inputenc}
\usepackage{etoolbox}
\usepackage{hhline}
\usepackage{array}
\usepackage{amssymb,amsmath,amsthm,amsfonts,graphicx}
\usepackage{latexsym}
\usepackage{bm}
\usepackage[]{hyperref}
\usepackage{cleveref}
\usepackage{nccmath}	%even more additional math symbols%

\makeatletter
    \patchcmd{\maketitle}{\@fpheader}{}{}{}
    \makeatother

\def\be{\begin{equation}}
\def\ee{\end{equation}}
\def\ba{\begin{eqnarray}}
\def\ea{\end{eqnarray}}

\title{Strong cosmic censorship for charged de Sitter black holes with a charged scalar field}

\author[a]{Oscar~J.~C.~Dias,}
\emailAdd{ojcd1r13@soton.ac.uk}
\affiliation[a]{STAG research centre and Mathematical Sciences, Highfield Campus, University of Southampton, Southampton SO17 1BJ, UK}

\author[b]{Harvey~S.~Reall}
\emailAdd{hsr1000@cam.ac.uk}
\affiliation[b]{Department of Applied Mathematics and Theoretical Physics, University of Cambridge, Wilberforce Road, Cambridge CB3 0WA, UK} 

\author[b]{and Jorge~E.~Santos}
\emailAdd{jss55@cam.ac.uk}

\abstract{
It has been shown recently that the strong cosmic censorship conjecture is violated by near-extremal Reissner-Nordstr\"om-de Sitter black holes. We investigate whether the introduction of a charged scalar field can rescue strong cosmic censorship. We find that such a field improves the situation but there is always a neighbourhood of extremality in which strong cosmic censorship is violated by perturbations arising from smooth initial data.
}

\begin{document}

\maketitle

%%%%%%%%%%%%%%%%%%%%%%%%%%%%%%%%%%%%%%%%%%%%%%%%%%%%%%%%%%%%
\section{Introduction}
%%%%%%%%%%%%%%%%%%%%%%%%%%%%%%%%%%%%%%%%%%%%%%%%%%%%%%%%%%%%

There has been recent interest in the strong cosmic censorship conjecture for theories with positive cosmological constant $\Lambda$. Cardoso {\it et al} have studied massless scalar field perturbations of Reissner-Nordstr\"om-de Sitter (RNdS) black holes \cite{Cardoso:2017soq}. Their results indicate that, when backreaction is included, a near-extremal black hole has a Cauchy horizon which is stable against perturbations in the sense that the perturbed spacetime can be extended across the Cauchy horizon as a weak solution of the equations of motion. This violates the version of strong cosmic censorship proposed by Christodoulou \cite{Christodoulou:2008nj}. However, generically, the extension across the Cauchy horizon will not be $C^2$ and so there is no violation of the $C^2$ version of strong cosmic censorship \cite{Cardoso:2017soq}.\footnote{We refer the reader to our recent paper \cite{Dias:2018etb} for a discussion of these different versions of strong cosmic censorship, and a summary of previous results on strong cosmic censorship with $\Lambda=0$ and $\Lambda>0$.}

These results are for Einstein-Maxwell theory coupled to a massless scalar. We have studied the analogous problem in pure Einstein-Maxwell theory, finding an even worse violation of strong cosmic censorship \cite{Dias:2018etb}. Our study of coupled linearized gravitational and electromagnetic perturbations of RNdS black holes indicates that, for a sufficiently large near-extremal black hole, perturbations can be extended across the Cauchy horizon in an arbitrarily smooth way. Thus not only are the Christodoulou and $C^2$ versions of strong cosmic censorship violated, but so is the $C^r$ version, for any $r \ge 2$.

Surprisingly, the situation for Kerr-de Sitter (Kerr-dS) black holes is completely different \cite{Dias:2018ynt}. In this case, massless scalar field and linearized gravitational perturbations both respect the Christodoulou version of strong cosmic censorship (and hence also the $C^2$ version). Thus there appears to be a qualitative difference between Einstein(-scalar field) theory and Einstein-Maxwell(-scalar field) theory as far as strong cosmic censorship is concerned. 

It is interesting to ask whether this difference might be related to another qualitative difference between the theories being considered here. In the RNdS case, there is no way of forming the black hole in Einstein-Maxwell theory or Einstein-Maxwell coupled to an uncharged scalar. By contrast, in the Kerr-dS case, one can form the black hole in Einstein gravity via collisions of gravitational waves. This motivates considering a theory in which RNdS can be formed via collapse, i.e., a theory containing charged matter.

The simplest such theory is Einstein-Maxwell theory coupled to a charged scalar field. In this paper, we will investigate whether or not strong cosmic censorship is respected by RNdS black holes in this theory. It has already been noted in Ref. \cite{Hod:2018dpx} that, at least in a certain region of RNdS parameter space, a scalar field with large charge and mass exhibits much less smooth behaviour at the Cauchy horizon of a near-extremal black hole than an uncharged scalar, which is promising for strong cosmic censorship. Nevertheless, we will show that, even with a charged scalar field, {\it there is always a violation of Christodoulou's version of strong cosmic censorship in a neighbourhood of extremality}. However, for physical values of the scalar field and black hole parameters, this neighbourhood is very small.

In this paper, when we discuss strong cosmic censorship, we are always referring to perturbations arising from {\it smooth} initial data. Ref. \cite{Dafermos:2018tha} has shown that if one allows {\it rough} initial data then Christodoulou's version of strong cosmic censorship is true for linear perturbations of RNdS black holes. More precisely: the solution at the Cauchy horizon is, generically, rougher than the initial data. Our previous paper \cite{Dias:2018etb} contains an extended discussion of this work. 

Finally, we note that the possibility of a violation of strong cosmic censorship with positive $\Lambda$ was first discussed long ago \cite{Mellor:1989ac} but is was eventually concluded that a violation (of the $C^2$ version) does not occur \cite{Brady:1998au}, which is in disagreement with our discussion above. This disagreement is explained in our previous paper \cite{Dias:2018etb}, where we show that the argument of Ref. \cite{Brady:1998au} applies only to initial data which is not $C^1$ at the event horizon. So the argument of Ref. \cite{Brady:1998au} is really a precursor of the rough version of strong cosmic censorship proposed in Ref. \cite{Dafermos:2018tha} and says nothing about what happens for smooth initial data.

This paper is organized as follows: in section \ref{sec:background} we will review the RNdS solution and explain why strong cosmic censorship for such black holes can be investigated by looking at quasinormal modes. In section \ref{sec:chargedScalarSCC} we discuss some general features of charged scalar quasinormal modes and how we calculate them. Section \ref{sec:smallq} presents some analytical results for the case of small scalar field charge, for which an instability occurs. In section \ref{sec:NHqnmcharged} we discuss analytical results for near-extremal black holes. Section \ref{sec:WKBqnmcharged} describes a WKB analysis of the case of large scalar field charge. Section \ref{sec:numerics} presents our numerical results. Finally, section \ref{sec:discussion} contains a brief discussion. 

%%%%%%%%%%%%%%%%%%%%%%%%%%%%%%%%%%%%%%%%%%%%%%%%%%%%%%%%%%%%
\section{Background}
\label{sec:background}

\subsection{\label{sec:gen}Reissner-Nordstr\"om de Sitter black holes }
%%%%%%%%%%%%%%%%%%%%
We will be investigating the quasinormal mode spectrum of charged perturbations around a Reissner-Nordstr\"om de Sitter (RNdS) black hole. This black hole is a solution of Einstein-Maxwell theory endowed with a positive cosmological constant $\Lambda \equiv 3/L^2$. The action is
\be
S=\frac{1}{16\pi G}\int \mathrm{d}^4 x \sqrt{-g}\left(R -\frac{6}{L^2}-F^2 \right)\,,
\ee
where $R$ is the Ricci scalar of the metric $g$ and $F=\mathrm{d}A$ is the Maxwell field strength associated to the potential 1-form $A$.

In static coordinates $(t,r,\theta,\phi)$, the RNdS solution with mass and charge parameters $M$ and $Q$ is 
\be
\mathrm{d}s^2=-f\,\mathrm{d}t^2+\frac{\mathrm{d}r^2}{f}+r^2\mathrm{d}\Omega_2^2\,,\qquad A=-\frac{Q}{r}\mathrm{d}t\,,
 \label{metricRN}
\ee
with $\mathrm{d}\Omega_2^2$ being the line element of a unit radius $S^2$ (parametrised by $\theta$ and $\phi$) and 
\be
f(r)=1-\frac{r^2}{L^2} -\frac{2M}{r}+\frac{Q^2}{r^2}\,.
\label{metricRNaux}
\ee

For an appropriate range of parameters, which we will specify below, the function $f$ has $3$ positive roots $r_-\le r_+\le r_c$ corresponding to the Cauchy horizon $\mathcal{CH}^+$, event horizon  $\mathcal{H}^+_R$  and cosmological horizon $\mathcal{H}_C^+$ respectively. We will denote the (positive) surface gravities associated to each of these three horizons as $\kappa_-,\kappa_+$ and $\kappa_c$, respectively. For any non-extremal RNdS black hole it can be shown that \cite{Brady:1998au}
\be
\label{kappaineq}
\kappa_- > \kappa_+ \,.
\ee 
The extremal configuration occurs when $\kappa_+=\kappa_-=0$. This happens when $Q=Q_{\rm ext}$ where
\be
Q_{\rm ext}=y_+ \, \sqrt{\frac{1+2\,y_+}{1+2\,y_+ + 3\,y_+^2}}\,r_c\,,\qquad \hbox{with}\quad y_+=\frac{r_+}{r_c}\,.
\label{Qext}
\ee
When presenting many of our results and associated plots we will parametrize  the RNdS solution using the dimensionless parameters $Q/Q_{\rm ext}$ and $y_+$.

\subsection{Quasinormal modes and strong cosmic censorship}

It was argued long ago that, for smooth initial data, the behaviour of generic linear perturbations at the Cauchy horizon of a RNdS black hole is determined by {\it quasinormal modes} of the black hole \cite{Mellor:1989ac}. See our previous work Ref. \cite{Dias:2018etb} for an extended review of this argument. This argument assumes that the initial perturbation vanishes at the bifurcation spheres of the event and cosmological horizons. However, this restriction has been eliminated by more recent work in the mathematics literature \cite{Hintz:2015jkj}. To state the conclusion, we define the {\it spectral gap} $\alpha$ as the minimum value of $-{\rm Im}(\omega)$ over all quasinormal frequencies $\omega$. We then define
\be
\label{betadef}
\beta = \frac{\alpha}{\kappa_-}\,.
\ee 
We assume that we are discussing scalar field perturbations. The conclusion is that if $\beta<1$ then the scalar field will, generically, fail to be $C^1$ at the ``right" (or ``ingoing") Cauchy horizon (i.e. the component of the future Cauchy horizon nearest to the future event horizon on the Penrose diagram). Conversely if $\beta>1$ then generic scalar field perturbations can be extended across the Cauchy horizon in $C^1$. However, in the Christodoulou version of strong cosmic censorship, the relevant question is whether the first derivative of the scalar field is locally square integrable at the Cauchy horizon.\footnote{A function is ``locally square integrable" if it is square integrable when multipled by any smooth test function of compact support.} Roughly speaking, this corresponds to the condition that the scalar field should have finite energy at the Cauchy horizon. See Refs. \cite{Cardoso:2017soq,Dias:2018etb} for discussion of the motivation for this condition. The condition for the scalar field generically {\it not} to have finite energy at the Cauchy horizon is
\be
\label{sccbound}
\beta < 1/2\,.
\ee
Conversely, if $\beta>1/2$ then scalar field perturbations arising from smooth initial data have finite energy at the Cauchy horizon. When gravitational backreaction is included, various nonlinear results \cite{Hintz:2016gwb,Hintz:2016jak,Costa:2017tjc} suggest if $\beta>1/2$ then Christodoulou's version of strong cosmic censorship will be violated whereas if $\beta<1/2$ then it will be respected. Thus (\ref{sccbound}) is regarded as the condition for Christodoulou's version of strong cosmic censorship to hold. 
 
Ref. \cite{Cardoso:2017soq} showed that, for a massless uncharged scalar field, near-extremal RNdS black holes have $1/2 < \beta < 1$. Thus Christodoulou's version of strong cosmic censorship is violated by such black holes. Ref. \cite{Dias:2018etb} showed that, for sufficiently large near-extremal RNdS black holes, linearized electromagnetic and gravitational perturbations can have {\it arbitrarily large} $\beta$ and so such perturbations can be arbitrarily smooth at the Cauchy horizon. But for Kerr-dS black holes, Ref. \cite{Dias:2018ynt} showed that \eqref{sccbound} is always respected by linearized gravitational (or massless scalar field) perturbations and so such black holes respect strong cosmic censorship. 

In the rest of this paper, we will investigate whether or not \eqref{sccbound} is respected by {\it charged} scalar field perturbations. 

%%%%%%%%%%%%%%%%%%%%%%%%%%%%%%%%%%%%%%%%%%%%%%%%%%%%%%%%%%%%
\section{\label{sec:chargedScalarSCC}Quasinormal modes}
%%%%%%%%%%%%%%%%%%%%

In this section we will analyse the quasinormal mode spectrum of a charged scalar perturbation of a RNdS black hole. Such perturbations are governed by the following complex linear equation
\begin{equation}
\mathcal{D}^2 \Phi -\mu^2 \Phi=0\,,
\end{equation}
where $\mathcal{D}=\nabla - i\,q\, A$, $q$ is the scalar field charge and $\mu$ its mass. Since the background (\ref{metricRN}) is static and spherically symmetry we can separate such perturbations as follows
\be
\Phi(t,r,\theta,\phi) = e^{-i\omega\, t}\,Y_{\ell m}(\theta,\phi)\,\widehat{\Phi}_{\omega\ell}(r)
\ee
where $Y_{\ell  m}(\theta,\phi)$ are the standard scalar harmonics on the unit round $S^2$. The quasinormal mode spectrum is then obtained by computing all the eigenpairs $(\omega,\widehat{\Phi}_{\omega\ell})$, for fixed $\{q\,r_c,\mu\,r_c,y_+,Q/Q_{\mathrm{ext}},\ell\}$\footnote{Note that we used the scaling symmetry $\{t,r,\theta,\phi\}\to\{\lambda t, \lambda r,\theta,\phi\}$ to measure everything in units of $r_c$.} subject to appropriate boundary conditions. The resulting equation for $\widehat{\Phi}_{\omega\ell}$ is of the generalized Sturm-Liouville type
\be
\left(f\,r^2 \widehat{\Phi}_{\omega \ell}^\prime\right)^\prime+\left[\frac{r^2}{f}\left(\omega-\frac{q\,Q}{r}\right)^2-\ell(\ell+1)-\mu^2\,r^2\right]\widehat{\Phi}_{\omega\ell}=0\,,
\label{eq:separ}
\ee
which reduces to the equation studied in \cite{Cardoso:2017soq} for $q=\mu=0$. We now introduce new coordinates adapted to our numerical scheme, namely
\be
y \equiv \frac{r-r_+}{r_c-r_+}\,,
\ee
where $r_+$ is the radius of the black hole, and $r_c$ the radius of the cosmological horizon. By construction, $y=1$ marks the location of the cosmological horizon, and $y=0$ the black hole horizon.

We now turn to the thorny issue of boundary conditions. We want to impose regularity across the future event horizon of the black hole, and at the cosmological horizon. This demands ingoing boundary conditions at $\mathcal{H}^+$ and outgoing at $\mathcal{H}^+_c$. To understand what this means for our scalar field $\widehat{\Phi}_{\omega \ell}(y)$, we perform a Frobenius analysis close to $y=0$, which gives
\be
\widehat{\Phi}_{\omega \ell}(y)\approx y^{\pm\,i\,\alpha_+}\left[1+\mathcal{O}(y)\right]\,,
\label{eq:lower1}
\ee
where
$$
\alpha_+ = \frac{y_+^2 \left(y_+^2+y_++1\right) \left(y_+ \tilde{\omega }-\tilde{q} \tilde{Q}\right)}{\left(1-y_+\right) \left[y_+^2 \left(2 y_++1\right)-\left(3 y_+^2+2 y_++1\right) \tilde{Q}^2\right]}=\frac{1}{2\kappa_+}\left(\tilde{\omega }-\frac{\tilde{q} \tilde{Q}}{y_+}\right)\,,
$$
with 
\be
  y_+\equiv r_+/r_c
\ee
and a tilde is used for other quantities measured in units of $r_c$:
\be
\tilde{Q}\equiv Q/r_c \qquad \tilde{\omega}\equiv\omega\,r_c  \qquad \tilde{q}\equiv q\,r_c \qquad \tilde{\mu}=\mu\,r_c\,.
\ee
Ingoing boundary conditions at the black hole horizon demands choosing the lower sign in \eqref{eq:lower1}. A similar analysis around $y=1$ reveals
\be
\widehat{\Phi}_{\omega \ell}(y)\approx (1-y)^{\pm\,i\,\alpha_c}\left[1+\mathcal{O}(1-y)\right]\,,
\label{eq:lower2}
\ee
where
$$
\alpha_c = \frac{y_+ \left(y_+^2+y_++1\right) \left(\tilde{\omega }-\tilde{q} \tilde{Q}\right)}{\left(1-y_+\right) \left[y_+ \left(y_++2\right)-\left(y_+^2+2 y_++3\right) \tilde{Q}^2\right]}=\frac{1}{2\kappa_c}\left(\tilde{\omega }-\tilde{q} \tilde{Q}\right)\,.
$$
Again, choosing outgoing boundary conditions at the cosmological horizon requires choosing the lower sign in \eqref{eq:lower2}.

Since we are going to use a Chebyshev collocation scheme to numerically solve for $(\omega,\tilde{\Phi}_{\omega\ell})$, we want to perform a change of variable such that the variable we solve for is a smooth function of $y$. This motivates the following redefinition
\be
\widehat{\Phi}_{\omega \ell}(y)= y^{-\,i\,\alpha_+}(1-y)^{-\,i\,\alpha_c}\mathcal{Q}_{\omega\ell}(y)\,,
\label{eq:red}
\ee
where, for our choice of boundary conditions, $\mathcal{Q}_{\omega\ell}(y)$ admits a regular Taylor series expansion at both $y=0$ and $y=1$. The boundary conditions for $\mathcal{Q}_{\omega\ell}(y)$ are then found to be of the Robin type, \emph{i.e.}
\be
\left.\partial_y\,\mathcal{Q}_{\omega\ell}(y)\right|_{y=0}=\mathfrak{G}(\tilde{\omega},\ell,\tilde{Q},\tilde{q},y_+,\tilde{\mu})\mathcal{Q}_{\omega\ell}(0)\,,
\ee
%A similar expression also holds for $y=1$, but with a different function $\mathfrak{G}$. 
Here, $\mathfrak{G}(\tilde{\omega},\ell,\tilde{Q},\tilde{q},y_+,\tilde{\mu})$ is a function that can be found by inserting \eqref{eq:red} into \eqref{eq:separ}, assuming a Taylor expansion for $\mathcal{Q}_{\omega\ell}(y)$ around $y=0$. A similar analysis can be done for $y=1$.

Note that if $\Phi$ has charge $q$ then $\Phi^*$ has charge $-q$. Furthermore, complex conjugation maps quasinormal modes to quasinormal modes. Therefore if $\omega=\omega_1 + i \omega_2$ is a quasinormal frequency of $\Phi$ then $-\omega^* = -\omega_1 + i \omega_2$ is a quasinormal frequency of $\Phi^*$. It follows that there is no loss of generality in assuming that $qQ>0$ in our analysis: results for $qQ<0$ are obtained simply by reversing the sign of the real part of the quasinormal frequencies. Note that when we calculate $\omega$ we have to allow both positive and negative values of $\omega_1$.  

We could attempt to perform an exhaustive study of the quasinormal mode spectrum as we did for gravitoelectromagnetic perturbations in Ref. \cite{Dias:2018etb}. However, the main point of our work is to show that, particularly for large charge $q$,  we can always find black hole solutions for which strong cosmic censorship is violated. This always seems to occur near extremality, so we shall focus our attention on near-extremal RNdS black holes. 

We will calculate quasinormal modes using a combination of analytical and numerical methods (the latter are explained and reviewed in \cite{Dias:2010eu,Cardoso:2013pza,Dias:2015nua}). There are three analytically tractable regimes that we will investigate: (i) the small $\tilde{q}$ regime, (ii) the near extremal limit, and (iii) the large $\tilde{q}$ limit. 

A physical value for the scalar field charge $q$ should be a multiple of the electron charge, which gives $|q| \sim 0.1$ in Planck units. In the real world, $r_c$ is enormous in Planck units and so in the physically interesting region of the RNdS parameter space we have $\tilde{q} \ggg 1$.  

%%%%%%%%%%%%%%%%
\section{\label{sec:smallq}Small scalar field charge: an instability}
%%%%%%%%%%%%%%%%

It has been shown previously that an instability can occur for small scalar field charge \cite{Zhu:2014sya,Konoplya:2014lha}. The earlier work discovered this instability using numerical methods. In this section we will study the instability analytically using perturbation theory. The regime of parameter space we will study is $\tilde{q} \ll 1$. As discussed above, this is not relevant physically but we will discuss it for completeness, and because we will later want to follow the behaviour of quasinormal modes as we increase $q$ starting with $q=0$. 

In this section we will assume that the scalar field is massless, i.e., $\mu=0$ and we will also set $\ell=0$. The quasinormal modes of interest become trivial, i.e., constant in spacetime, as $q \rightarrow 0$. Recall that the $\ell=\omega=0$ mode of a neutral massless scalar wave equation admits has a global shift symmetry of the form $\widehat{\Phi}_{0\,0}\to\widehat{\Phi}_{0\,0}+c$, where $c$ is a constant. This mode does not couple to gravity, thus carrying no energy, since in this limit only radial derivatives appear in the scalar stress energy tensor. However, when $\tilde{q}\neq0$, the mode becomes physical since the covariant derivatives appearing in the charged scalar stress energy tensor become non-vanishing. The triviality at $q=0$ is the reason why this mode is amenable to an analytic treatment.

We take an expansion of the following form
\begin{subequations}
\begin{align}
&\widehat{\Phi}_{\omega\,0}(r)=\left(\frac{r}{r_+}-1\right)^{-i\alpha_+}\left(1-\frac{r}{r_c}\right)^{-i\alpha_c} \sum_{n=0}^{+\infty} \phi^{(n)}_{\omega\,0}(r)\tilde{q}^n+\text{non-perturbative}\,,
\\
& \omega = \sum_{n=0}^{+\infty} \omega^{(n)}\tilde{q}^n+\text{non-perturbative}\,,
\end{align}
\label{eq:perturbsmallq}
\end{subequations}
where ``non-perturbative" refers to terms that are non-perturbative in $\tilde{q}$. We solve the equations in a power series, demanding regularity at the black hole and cosmological horizons. As a normalisation, we choose $\phi^{(0)}_{\omega\ell}(r_+)=1$ and $\phi^{(n)}_{\omega\ell}(r_+)=0$ for $n\geq1$.

When $\tilde{q}=0$, we know what the solutions look like, namely the trivial solution reads
\be
\phi^{(0)}_{\omega\ell}(r)=1\,\quad\text{and}\quad \omega^{(0)}=0\,.
\ee

One can now use this information, and move to the next order. Things become more complicated, but a solution for $\phi^{(1)}_{\omega\ell}(r)$ can still be found in closed form, and requiring regularity at $r=r_+$ and $r=r_c$ gives
\be
\label{omega1}
\omega^{(1)}=\frac{Q \left(r_c+r_+\right)}{r_c \left(r_c^2+r_+^2\right)}\,.
\ee
The second order calculation is substantially more complicated. While a solution for $\phi^{(2)}_{\omega\ell}(r)$ can be found in terms of Polylog functions, it is manifestly unwieldy to impose regularity at $r=r_+$ or $r=r_c$. Here, we proceed in a different manner more analogous to standard perturbation theory which was used with success in \cite{Cardoso:2013pza}. The idea is to use the fact that we know $\phi^{(0)}_{\omega\ell}(r)$ and $\phi^{(1)}_{\omega\ell}(r)$ to compute $\omega^{(2)}$. We start by multiplying Eq.~(\ref{eq:separ}) by $w(r)\phi^{(0)}_{\omega\ell}$, and using $\Phi_{\omega\ell}$ and $\omega$ given as in Eqs.~(\ref{eq:perturbsmallq}). Schematically, to order $q^2$, this leads to the following type of differential equation in $\phi^{(2)}_{\omega\ell}(r)$
\be
w(r)\phi^{(0)}_{\omega\ell}(r)\mathcal{D}_2\phi^{(0)}_{\omega\ell}(r)+w(r)\phi^{(0)}_{\omega\ell}(r)\mathcal{D}_1\phi^{(1)}_{\omega\ell}(r)+w(r)\phi^{(0)}_{\omega\ell}(r)\mathcal{D}_0\phi^{(2)}_{\omega\ell}(r)=0\,,
\ee
where $\mathcal{D}_i$ are, at most, second order differential operators in $r$, with $\mathcal{D}_2$ depending explicitly on $\omega^{(2)}$. Up to this point, $w(r)$ is arbitrary. We then integrate the above equation in $r$ between $r=r_+$ and $r=r_c$, leading to
\be
\int_{r_+}^{r_c}\mathrm{d}r\left[w(r)\phi^{(0)}_{\omega\ell}(r)\mathcal{D}_2\phi^{(0)}_{\omega\ell}(r)+w(r)\phi^{(0)}_{\omega\ell}(r)\mathcal{D}_1\phi^{(1)}_{\omega\ell}(r)+w(r)\phi^{(0)}_{\omega\ell}(r)\mathcal{D}_0\phi^{(2)}_{\omega\ell}(r)\right]=0\,.
\ee
Finally, we integrate the last term by parts, removing all derivatives from $\phi^{(2)}_{\omega\ell}(r)$ and choose $w(r)$ appropriately so that the last term vanishes\footnote{In doing this one generates boundary terms at the cosmological and black hole horizon, but they can be shown to vanish.}. This occurs for $w(r)=r^2$. The final equality only depends on $\phi^{(0)}_{\omega\ell}(r)$, $\phi^{(1)}_{\omega\ell}(r)$ and $\omega^{(2)}$, with the first two functions being known analytically. After integrating, one can then solve for $\omega^{(2)}$. We have done this exercise and found
\begin{subequations}
\begin{multline}
r_c\,\omega^{(2)}=-i\,\frac{y_+ \left(1-y_+^3\right) \tilde{Q}^2}{\left(1+y_+^2\right){}^3 \left(y_+-\tilde{Q}^2\right)}\Bigg\{\frac{\left(1+y_+^2\right) \left(1+y_+\right) }{1-y_+}\log \left(\frac{\kappa _+}{\kappa _c}y_+^2\right)+\left(1+y_+\right)^2
\\
+\frac{2 y_+ \left[1+y_++4 y_+^2+y_+^3+y_+^4+2 \left(1+y_+^2\right) \tilde{Q}^2\right]}{y_++\left(4+y_+\right) y_+^2-3 \left(1+y_+\right)^2 \tilde{Q}^2}\frac{\mathrm{arctanh}\;X(y_+,\tilde{Q})}{X(y_+,\tilde{Q})}\Bigg\}\,,
\label{eq:secondorder}
\end{multline}
where\footnote{It can be shown that $X$ is real across the whole RNdS parameter space and that $0\leq X\leq1$ with equality attained only at boundaries of the parameter space.}
\be
X(y_+,\tilde{Q})\equiv \frac{\left(1-y_+\right) \sqrt{y_+-\tilde{Q}^2} \sqrt{y_+\left(1+y_+\right)^2+\left(3+2\,y_++3 y_+^2\right) \tilde{Q}^2}}{y_++\left(4+y_+\right) y_+^2-3 \left(1+y_+\right)^2 \tilde{Q}^2}\,.
\ee
\end{subequations}

Some comments are in order regarding Eq.~(\ref{eq:secondorder}). First, it is purely imaginary, and its sign determines whether an instability exists for arbitrarily small values of $\tilde{q}$. Second, $|r_c\,\omega^{(2)}|$ diverges logarithmically as we approach extremality. In fact, we have
\begin{equation}
\left.r_c\,\omega^{(2)}\right|_{\tilde{Q}\simeq Q_{\mathrm{ext}/r_c}}\approx \frac{2 i y_+^3 \left(1+2 y_+\right) \left(1-2 y_+-y_+^2\right)}{\left(1+3 y_+\right) \left(1+y_+^2\right)^3}\left[-\log \left(\frac{Q_{\text{ext}}-Q}{r_c}\right)\right]\,.
\end{equation}
If $y_+ < \sqrt{2}-1$ then the RHS has positive imaginary part and so there is an instability if we are sufficiently close to extremality. Conversely, if $y_+> \sqrt{2}-1$ the the RHS has negative imaginary part, and an instability is not present near extremality.

\begin{figure}[ht]
	\centering
	\includegraphics[width=0.63\textwidth]{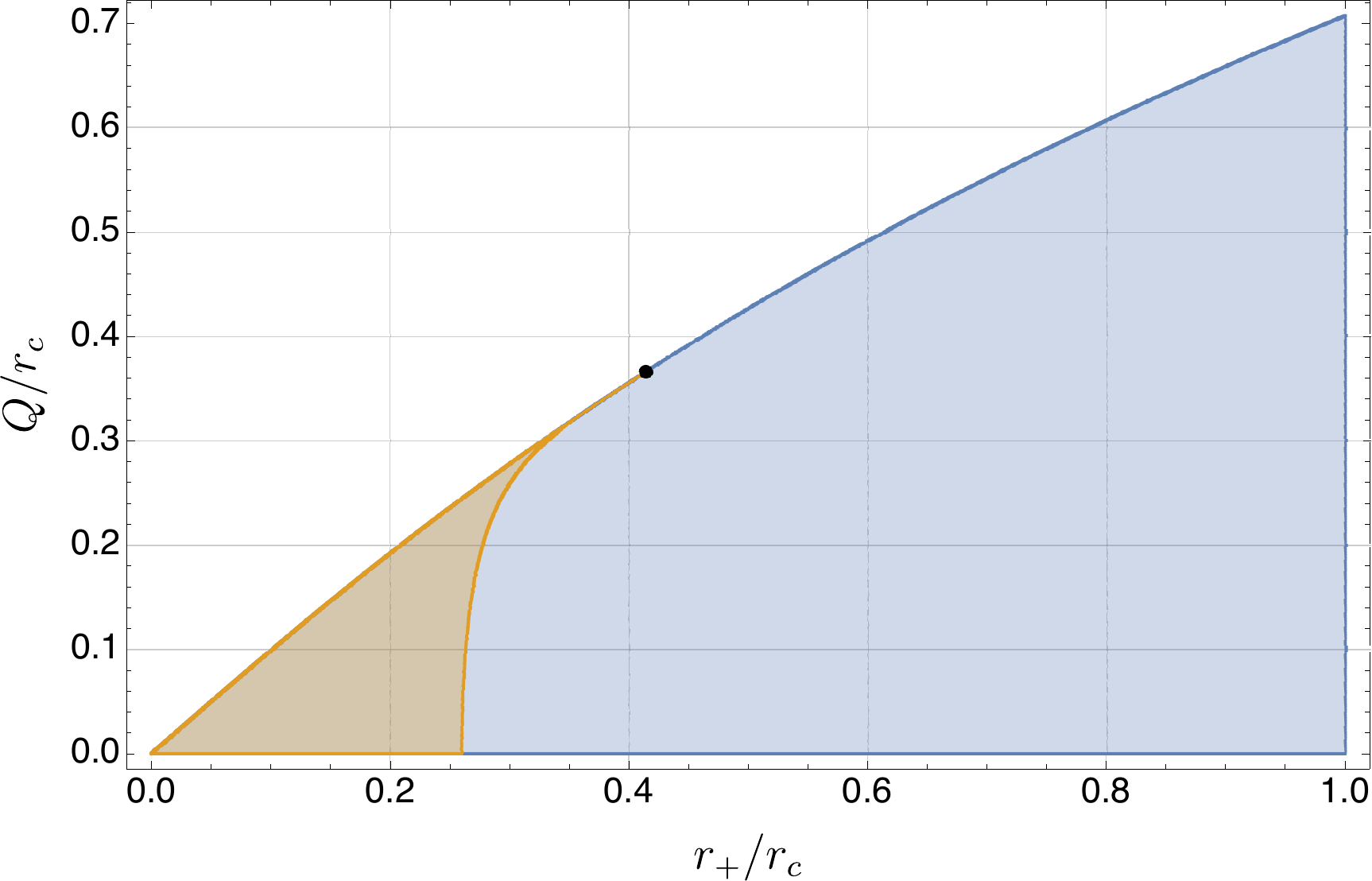}
	\caption{Moduli space of RNdS black holes: the orange region indicates the moduli space of RNdS black holes for which our perturbative calculations shows that an instability exists for arbitrarily small scalar field charge $\tilde{q}$ (and vanishing mass $\mu$), while the blue region is where the perturbative calculation indicates stability. The black dot corresponds to an extremal black hole with $y_+=\sqrt{2}-1$.}
	\label{fig:moduli}
\end{figure}

We can use \eqref{eq:secondorder} to map out the region of moduli space of RNdS black hole for which our perturbative calculation indicates that there is an instability for small $\tilde{q}$. This is shown in Fig.~\ref{fig:moduli}. Our numerical results below will show that, as we increase $\tilde{q}$, any instabilities continue to lie in the orange region of Fig.~\ref{fig:moduli}.
% the reason being that all instabilities we find smoothly connect to $\tilde{q}\to0$.

Note that if $y_+> \sqrt{2}-1$,then our perturbative calculation shows that these modes will  have $-\mathrm{Im}(\omega/\kappa_-)\to+\infty$ as we approach extremality, since $r_c\,\omega^{(2)}$ does not vanish at extremality. We shall see that this perturbative result remains true at finite $\tilde{q}$.
%%%%%%%%%%%%%%%%
\section{\label{sec:NHqnmcharged}Near-extremal family of charged scalar modes}
%%%%%%%%%%%%%%%%

Refs. \cite{Cardoso:2017soq,Dias:2018etb} found that strong cosmic censorship was violated by near-extremal RNdS black holes. Furthermore, these references found that, for such black holes, the most slowly decaying quasinormal modes were ``near-extremal" modes localized near the event horizon. These modes are amenable to an analytical treatment which we now discuss for the case of a charged scalar field. 

We introduce the following dimensionless variables
\be
\label{NHregime2}
x=1-\frac{r}{r_+}\,,\qquad\text{and}\qquad \sigma\equiv1-\frac{r_-}{r_+}\,.
\ee
We now consider a function redefinition of the following form
\be
\widehat{\Phi}_{\omega \ell}(r)=\hat{f}\left(\frac{x}{\sigma}\right)\,,
\ee
and expand \eqref{eq:separ} to first order in $\sigma$ while keeping $x/\sigma = z$ fixed and setting $\omega = Q q/r_+\,+\,\sigma\,\widetilde{\delta \omega}/r_c$. This ensures that we are zooming near $x=0$ as we take the near extremal limit (\emph{i.e.} zoom around $x\sim 0$). The equation simplifies substantially in this limit
\be
(1-z)z\frac{\mathrm{d}^2}{\mathrm{d}z^2}\hat{f}(z)+(1-2\,z)\frac{\mathrm{d}}{\mathrm{d}z}\hat{f}(z)+\left[\frac{\hat{\varphi}^2-\hat{\lambda}\, z}{z(1-z)}+\hat{\eta}\right]\hat{f}(z)=0\,,
\label{eq:nearHcs}
\ee
where we defined
\begin{align}
&\hat{\varphi}=\frac{y_+ \left(1+2y_+ +3 y_+^2\right)}{\left(1-y_+\right) \left(1+3 y_+\right)}\widetilde{\delta \omega}\,,\nonumber
\\
&\hat{\lambda}=\frac{y_+^2 \left(1+2\,y_++3\,y_+^2\right) \tilde{q}}{\left(1-y_+\right)^2(1+3y_+)^2}\left(2 \sqrt{1+4 y_++7 y_+^2+6 y_+^3}\;\widetilde{\delta \omega }-2 y_+ \tilde{q}-\tilde{q}\right)\,,\nonumber
\\
& \hat{\eta}=\frac{1+2\,y_++3\,y_+^2}{(1-y_+)(1+3y_+)}\left[y_+^2 \tilde{\mu}^2+\ell  (\ell +1)-\frac{\left(2 y_++1\right) y_+^2}{\left(1-y_+\right) \left(1+3 y_+\right)} \tilde{q}^2\right]\,.
\end{align}
Note that $\hat{\varphi}$ and $\hat{\lambda}$ depend on $\widetilde{\delta \omega}$, but $\hat{\eta}$ does not.

Equation (\ref{eq:nearHcs}) can be readily solved in terms of Gaussian Hypergeometric functions ${}_2F_1$. There are two linearly independent solutions to this equation, but only one of them is regular at the black hole event horizon, namely
\be
\hat{f}(z)=\hat{C}\,z^{-i\,\hat{\varphi}}(1-z)^{i\sqrt{\hat{\varphi}^2-\hat{\lambda}}}{}_2F_1\left(a_-,a_+\,;\,1-2\,i\,\hat{\varphi}\,;\,z\right)\,,
\label{eq:hypergcs}
\ee
where
\be
a_{\pm}=\frac{1}{2}\pm\sqrt{\frac{1}{4}+\hat{\eta}}+i\left(\sqrt{\hat{\varphi}^2-\hat{\lambda}}-\hat{\varphi}\right)\,.
\ee
and $\hat{C}$ is a constant. Ideally, we would like to match the behaviour of this solution to an outer solution that is outgoing at the cosmological horizon, i.e., perform a matched asymptotic expansion. However, we were not able to find an outer solution analytically. We thus proceed in a different manner. We will expand the Gaussian Hypergeometric function appearing in \eqref{eq:hypergcs} at large negative values of $z$ (equivalently, large positive values of $r$) and impose some rather \emph{ad hoc} boundary conditions there which can be physically motivated and that will partially match our numerical results. We stress that without having the outer solution, we are not expecting this procedure to give a very accurate approximation to the exact results. Nevertheless, we will see later that it does help understand some aspects of our numerical results. 

At large negative values of $z$, we get
\begin{multline}
\hat{f}(z)\approx -\frac{e^{-\pi\hat{\varphi}}}{\sqrt{-z}}\hat{C}\;\Gamma(1-2i\hat{\varphi})\times
\\
\Bigg\{\frac{(-1)^{\sqrt{1+4\hat{\eta}}}\Gamma\left(\sqrt{1+4\hat{\eta}}\right)}{\Gamma\left(a_+\right)\Gamma\left(b_+\right)}(-z)^{\frac{1}{2}\sqrt{1+4\hat{\eta}}}\left[1-\frac{\hat{\eta}-\hat{\lambda}-(\hat{\eta}+\hat{\lambda})\sqrt{1+4\hat{\eta}}}{4 \hat{\eta} }\frac{1}{(-z)}+\mathcal{O}(z^{-2})\right]+
\\
\frac{(-1)^{-\sqrt{1+4\hat{\eta}}}\Gamma\left(-\sqrt{1+4\hat{\eta}}\right)}{\Gamma\left(a_-\right)\Gamma\left(b_-\right)}(-z)^{-\frac{1}{2}\sqrt{1+4\hat{\eta}}}\left[1-\frac{\hat{\eta}-\hat{\lambda}+(\hat{\eta}+\hat{\lambda})\sqrt{1+4\hat{\eta}}}{4 \hat{\eta} }\frac{1}{(-z)}+\mathcal{O}(z^{-2})\right]
\Bigg\}\,,
\label{eq:large}
\end{multline}
where
\be
b_{\pm}=a_{\pm}-2i\sqrt{\hat{\varphi}^2-\hat{\lambda}}\,.
\ee
Our boundary conditions will depend on whether $1+4\hat{\eta}\in\mathbb{R}$ is positive or negative. For positive values of $1+4\hat{\eta}$, one term above grows at large $z$ and the other decays, and so we will choose our perturbation to vanish asymptotically, which means we want to kill the term proportional to $(-z)^{\frac{1}{2}\sqrt{1+4\tilde{\eta}}}$. This boundary condition is motivated by the fact that the modes we are interested in are near horizon modes, so we want our solution to only have support near the black hole horizon. This boundary condition implies 
\be
a_+-2i\sqrt{\hat{\varphi}^2-\hat{\lambda}}=-p\,,
\ee
where $p\in\mathbb{N}_0=\{0,1,2,\ldots\}$. This equation can be readily solved for $\widehat{\delta\omega}$
\begin{multline}
\widehat{\delta\omega}=\widehat{\delta\omega}^{-}\equiv\frac{1}{4(1+2\,y_++3\,y_+^2)}\Bigg[2\tilde{q}\sqrt{1+4 y_++7 y_+^2+6 y_+^3}
\\-i\frac{(1-y_+)(1+3y_+)}{y_+}\left(1+2p+\sqrt{1+4\hat{\eta}}\right)\Bigg]\,.
\label{eq:deltaomegaminus}
\end{multline}
Note that this quantisation will be valid so long as $1+4\hat{\eta}>0$, or equivalently if
\be
|\tilde{q}|\leq\tilde{q}_c\equiv \frac{\left(1-y_+\right) \left(1+3 y_+\right)}{2 y_+ \sqrt{\left(1+2 y_+\right) \left(1+2 y_++3 y_+^2\right)}}\sqrt{1+\frac{4 \left(1+2 y_++3 y_+^2\right) \left[y_+^2 \tilde{\mu}^2+\ell  (\ell +1)\right]}{\left(1-y_+\right) \left(1+3 y_+\right)}}\,,
\label{eq:qc}
\ee
We will also define $q_c = \tilde{q}_c/r_c$. 

It is interesting to take $q$ to be a multiple of the electron charge, so $|q| \sim 0.1$ in Planck units, and consider a (near-extremal) black hole for which $y_+$ is not close to $1$. Consider first the case in which the Compton wavelength of the field is small compared to the size of the black hole: $\mu r_+ \gg 1$, so $\tilde{\mu} y_+ \gg 1$. We then have $\tilde{q}_c \sim \tilde{\mu}$ so $q_c \sim \mu$. So we have $|q| \gg q_c$ if $\mu$ is well below the Planck mass. Now consider the opposite limit $\mu r_+ \ll 1$. This gives $q_c \sim 1/r_+$ so again we have $q \gg q_c$ if the black hole is large compared to the Planck length. So it appears that, for physically interesting values of the scalar field and black hole parameters, we will always have $ |q|  \gg q_c$ i.e. $1+4\hat{\eta}<0$.

So let us now discuss the case $1+4\hat{\eta}<0$. In this case both terms in \eqref{eq:large} oscillate rather than grow or decay. Our previous \emph{ansatz} is less motivated, so we need to consider a different set of boundary conditions. We will require these travelling waves to be purely outgoing with respect the phase velocity. This quantity can be computed using the machinery developed in \cite{Bardeen:1999px} and reviewed in great detail in the Appendix B of \cite{Dias:2009ex}.

The behaviour of our wavefunctions near $r\to+\infty$ can be encapsulated in the following function
\be
S_{\pm}(r)=\mathrm{exp}\left[\pm \frac{i}{2}\sqrt{\left|1+4\hat{\eta}\right|}\log\left(\frac{r}{r_+ \sigma}\right)-\frac{\hat{\eta}-\hat{\lambda}\mp(\hat{\eta}+\hat{\lambda})i\sqrt{\left|1+4\hat{\eta}\right|}}{4 \hat{\eta} }\frac{r_+\sigma}{r}+\mathcal{O}(r^{-2})\right]\,,
\ee
and where the upper (lower) sign corresponds to the first (last) term in the expansion (\ref{eq:large}). With this function we can define an effective $r$-dependent wavenumber via
\be
k_{\pm}(r)=-i\frac{1}{ S_{\pm}}\frac{\mathrm{d}S_{\pm}(r)}{\mathrm{d}r}\,,
\ee
which in turn induce the following $r$-dependent phase velocity 
\be
v_{\mathrm{ph}}^{\pm}(r)=\frac{\omega}{k_{\pm}(r)}\,.
\ee
At large $r$, we find
\be
v_{\mathrm{ph}}^{\pm}(r)\approx \pm\frac{2\,r\,\mathrm{Re}(\omega)  }{\sqrt{\left| 1+4 \eta \right| }}\,.
\ee
An outgoing phase velocity means choosing the term which has positive phase velocity for positive $\mathrm{Re}(\omega)$. This means we want to make the term in \eqref{eq:large} proportional to $(-z)^{-\frac{1}{2}\sqrt{1+4\tilde{\eta}}}$ vanish. We can achieve this setting
\be
a_--2i\sqrt{\hat{\varphi}^2-\hat{\lambda}}=-p\,,
\ee
where $p\in\mathbb{N}_0$. Just as above, we can solve this equation with respect to $\widehat{\delta\omega}$
\begin{multline}
\widehat{\delta\omega}=\widehat{\delta\omega}^{+}\equiv\frac{1}{4(1+2\,y_++3\,y_+^2)}\Bigg[2\tilde{q}\sqrt{1+4 y_++7 y_+^2+6 y_+^3}-\\ \frac{(1-y_+)(1+3y_+)}{y_+}\sqrt{\left|1+4\hat{\eta}\right|}-i\frac{(1-y_+)(1+3y_+)}{y_+}\left(1+2p\right)\Bigg]\,.
\label{eq:deltaomegaplus}
\end{multline}

To summarise, we have found
\begin{equation}
\widehat{\delta\omega}=\left\{
\begin{array}{c}
\widehat{\delta\omega}^{-}\quad \text{for}\quad |q|\leq q_c
\\
\,
\\
\widehat{\delta\omega}^{+}\quad \text{for}\quad |q|\geq q_c
\end{array}
\right.\,,
\label{eq:nhsum}
\end{equation}
with $\widetilde{\delta\omega}^{-}$ defined in \eqref{eq:deltaomegaminus}, $\widetilde{\delta\omega}^{+}$ defined in \eqref{eq:deltaomegaplus} and $q_c$ defined in \eqref{eq:qc}. We will compare the above analytical results to our numerical results below. 

%%%%%%%%%%%%%%%%
\section{\label{sec:WKBqnmcharged}Large scalar field charge: WKB analysis}
%%%%%%%%%%%%%%%%
Another analytically tractable limit of \eqref{eq:separ} is the WKB limit in which we take the charge $\tilde{q}$ to be large compared to $\mu\,r_c$ and $\ell$. One could preform a geodesic analysis similar to the one developed in \cite{Dias:2018etb}. However, due to the simplicity of the equation (\ref{eq:separ}) governing charged scalar perturbations, we aim to do better. Here we will follow \cite{Dolan:2010wr} \emph{mutatis mutandis}. We start by making an \emph{Ansatz} for the field perturbations, which we solve order by order in $1/q$. In particular, we postulate the following expansion in $1/q$
\begin{subequations}
\begin{align}
&\widehat{\Phi}_{\omega\ell}(r)=\left(\frac{r}{r_+}-1\right)^{-i\alpha_+}\left(1-\frac{r}{r_c}\right)^{-i\alpha_c} e^{-\,q\,\psi_{\omega\ell}(r)}\sum_{n=0}^{+\infty} \frac{\phi^{(n)}_{\omega\ell}(r)}{q^n}+\text{non-perturbative}\,,
\\
& \omega = \sum_{n=-1}^{+\infty} \frac{\omega^{(n)}}{q^n}+\text{non-perturbative}\,,
\label{eq:perturbwkb}
\end{align}
\end{subequations}
where ``non-perturbative" refers to terms that cannot be expanded as power series in $1/q$. 
We will later see that these non-perturbative terms are essential for deciding whether or not strong cosmic censorship is preserved in the presence of a charged scalar field. To leading order we find \emph{two} possibilities
\begin{subequations}
\begin{equation}
\omega^{(-1)}_+=\frac{Q}{r_+}\,,
\end{equation}
or
\begin{equation}
\omega^{(-1)}_c=\frac{Q}{r_c}\,.
\end{equation}
\end{subequations}
Each of these possibilities leads to different equations governing the behaviour of $\psi_{\omega\ell}(r)$
\begin{subequations}
\begin{align}
&\frac{\mathrm{d}\psi^+_{\omega\ell}(r)}{\mathrm{d}r}=-\frac{i}{2 f(r)\kappa_c}\left(1-\frac{r}{r_c}\right)\left(\frac{r}{r_+}-1\right)\frac{Q}{r}\left(\frac{1}{r_+}-\frac{1}{r_c}\right)\left(\frac{r_+ r_c-Q^2}{r_c^2+r_c\,r_++r_+^2}+\frac{Q^2}{r\,r_c}\right)\,,
\\
&\frac{\mathrm{d}\psi^c_{\omega\ell}(r)}{\mathrm{d}r}=\frac{i}{2 f(r)\kappa_+}\left(1-\frac{r}{r_c}\right)\left(\frac{r}{r_+}-1\right)\frac{Q}{r}\left(\frac{1}{r_+}-\frac{1}{r_c}\right)\left(\frac{r_+ r_c-Q^2}{r_c^2+r_c\,r_++r_+^2}+\frac{Q^2}{r\,r_+}\right)\,.
\end{align}
\end{subequations}
Unlike the uncharged case of \cite{Cardoso:2017soq}, we see that the WKB expansion now predicts two distinct families of quasinormal modes as $q$ gets large. We shall shortly see that these connect continuously to modes with $q=0$. We will call the first family the black hole family, and the second family the cosmological family and label quantities referring to either of the families with a subscript $+$ and $c$, respectively.

The expressions for $\phi^{(n)}_{\omega\ell}(r)$ are increasingly complicated at higher order, so we will omit them here. However, the first few corrections to $\omega$ can be written in a rather compact manner. For the black hole family we find
\begin{subequations}
\begin{align}
&\omega^{(0)}_+=-\frac{i}{2}\kappa_+\,,
\\
&\omega^{(1)}_+=\frac{\kappa_+}{2Q}\left\{r_+^2\mu^2+\ell(\ell+1)+\frac{1}{4}\left[9\frac{r_c^3+(Q^2+r_c^2)r_+}{r_c(r_c^2+r_c\,r_++r_+^2)}-7-\frac{Q^2}{r_+^2}\right]\right\}\,,
\\
& \omega^{(2)}_+=\frac{i\,\kappa_+^2}{16 Q^2 r_c\,r_+}\left[r_c\left(8 r_+^4\mu^2+3 Q^2-r_+^2\right)+15 \frac{r_+^3\left(Q^2-r_c\,r_+\right)}{r_c^2+r_c\,r_++r_+^2}\right]\,.
\end{align}
\label{eqs:wkbs}
\end{subequations}
whereas for the cosmological family we have
\begin{subequations}
\begin{align}
&\omega^{(0)}_c=-\frac{i}{2}\kappa_c\,,
\\
&\omega^{(1)}_c=-\frac{\kappa_c}{2Q}\left\{r_c^2\mu^2+\ell(\ell+1)+\frac{1}{4}\left[9\frac{r_+^3+(Q^2+r_+^2)r_c}{r_+(r_c^2+r_c\,r_++r_+^2)}-7-\frac{Q^2}{r_c^2}\right]\right\}\,,
\\
& \omega^{(2)}_c=-\frac{i\,\kappa_c^2}{16 Q^2 r_c\,r_+}\left[r_+\left(8 r_c^4\mu^2+3 Q^2-r_c^2\right)+15 \frac{r_c^3\left(Q^2-r_c\,r_+\right)}{r_c^2+r_c\,r_++r_+^2}\right]\,.
\end{align}
\label{eqs:wkbsc}
\end{subequations}

\begin{figure}[ht]
	\centering
	\includegraphics[width=0.45\textwidth]{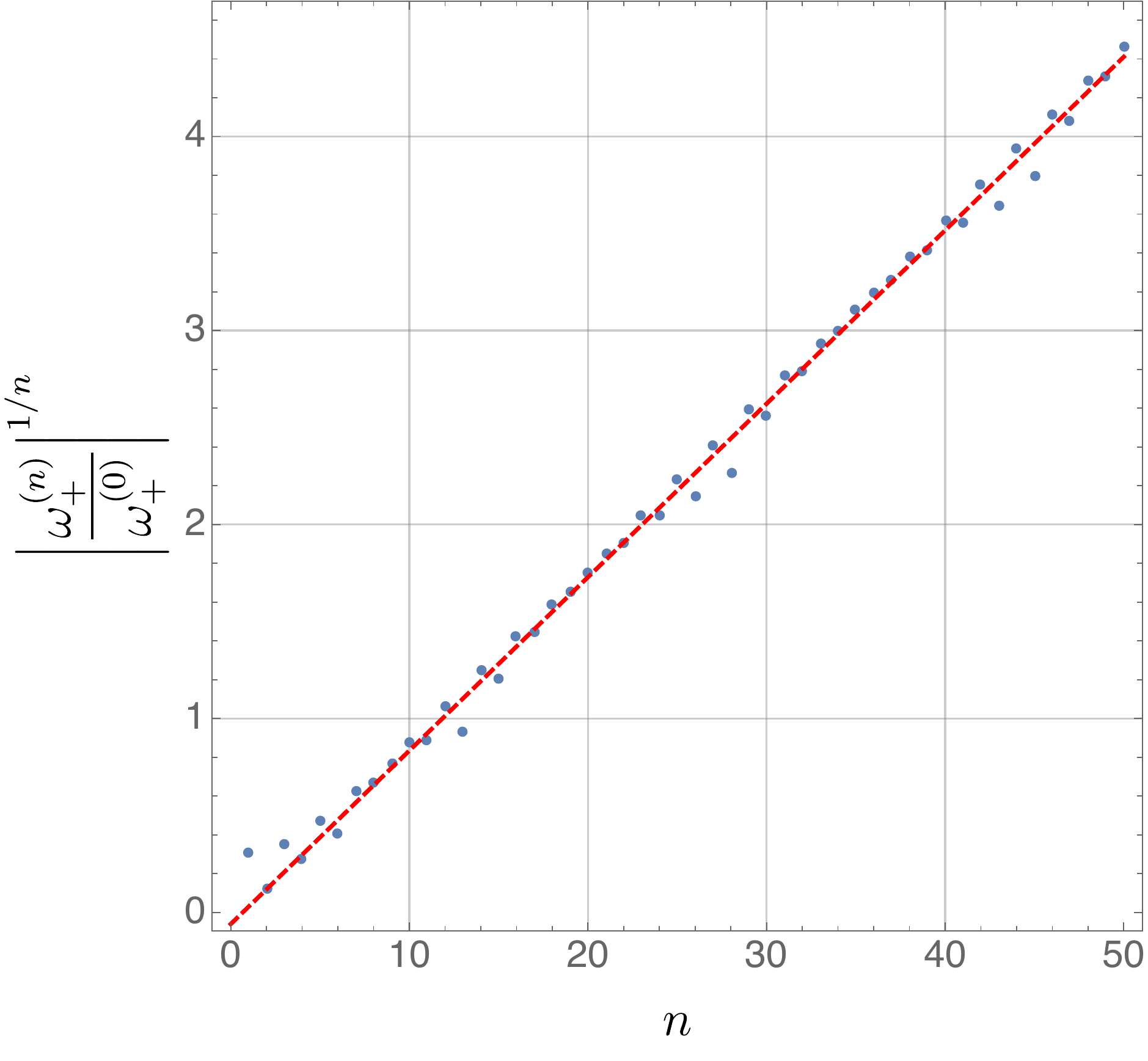}
	\caption{Coefficients of $\omega_+$ in the WKB expansion (\ref{eq:perturbwkb}) as a function of the WKB order $n$. The red dashed line shows a linear fit of the form $-0.059708(1) + 0.089413(2)\,n$, indicating consistency with factorial growth at large $n$. This particular curve was extracted for $r_c=1$, $r_+=1/2$, $\mu=0$, $Q=2/5$ and $\ell=0$.}
	\label{fig:wkb}
\end{figure}

For particular values of the parameters we have managed to analytically extend our calculation up to order $1/q^{50}$. This was rather informative since it allowed us to confirm some of the patterns emerging in the coefficients above. For instance, we find that all odd (even) orders in $1/q$ contribute to the real  (imaginary) part only. Furthermore, one can plot the magnitude of the corrections to the frequency as a function of the order $n$ as in Fig.~\ref{fig:wkb}. This particular curve was extracted for $r_c=1$, $r_+=1/2$, $\mu=0$, $Q=2/5$, $\ell=0$ and for the black hole family. Similar results hold for the cosmological family. The growth at large $n$ is consistent with factorial, indicating that, perhaps as expected, the WKB expansion is asymptotic only. To confirm this, we have performed a fit to $|\omega^{(n)}_+/\omega^{(0)}_+|^{1/n}$ for $n\geq8$, and found very good agreement with linear behaviour at large $n$ (this is depicted in Fig.~\ref{fig:wkb} as the dashed red curve).

We can now discuss strong cosmic censorship. The decay rates of the cosmological family remain non-zero as we approach extremality, whereas the decay rates of the black hole family approach zero as we approach extremality. Hence it is the black hole family which is relevant for calculating $\beta$ in \eqref{betadef} near extremality. From \eqref{sccbound} it follows that if we can show that there exists a quasinormal mode with $-{\rm Im}(\omega)/\kappa_-<1/2$ then strong cosmic censorship is {\it not} violated. Now, to leading order in $1/q$, and near extremality, the WKB expansion for the black hole family predicts
\be
-\frac{{\rm Im}(\omega_+)}{\kappa_-} = \frac{1}{2}-\frac{r_c^2+2r_c\,r_+-r_+^2}{(r_c-r_+)(r_c+3\,r_+)}\sigma+\mathcal{O}(\sigma^2)+\mathcal{O}(1/q)+\text{non-perturbative terms}\,.
\ee
Since the second term is negative definite, one might be tempted to conjecture that near extremality, strong cosmic censorship is respected. However, we note that this negative definite term vanishes as $\sigma\to0$, and so near-extremality one cannot neglect the non-perturbative terms. We will show numerically that the terms non-perturbative in $1/q$ do \emph{not} vanish near extremality. 

Stating things differently: for any given RNdS black hole, if one takes $q$ large enough then the non-perturbative terms are negligible and so the above result implies $\beta < 1/2$ for $q$ above some critical value $q_\star$. But $q_\star$ depends on the black hole parameters, and diverges in the extremal limit. Therefore, for any fixed value of $q$ there will exist near-extremal black holes for which the non-perturbative terms are significant and so one cannot conclude from the above expression that strong cosmic censorship is respected. The information about the non-perturbative terms might already be contained in the perturbative expansion (\ref{eq:perturbwkb}) via a clever Borel resummation of the WKB expansion outlined above, but we leave such an endeavour for future investigations.
%%%%%%%%%%%%%%%%%%%%%%%%%%%%%%%%
\section{Numerical results}
\label{sec:numerics}
%%%%%%%%%%%%%%%%%%%%%%%%%%%%%%%%
\subsection{\label{sec:general}Tracking the special mode}
%%%%%%%%%%%%%%%%%%%%%%%%%%%%%%%%

In this section we will study numerically the special mode that we discussed in section \ref{sec:smallq}. Recall that this is the mode with $\ell=0$ and $\mu=0$ (i.e. a massless field) which reduce to the trivial (constant) mode in the limit $q \rightarrow 0$. The perturbative analysis of section \ref{sec:smallq} is valid for small $\tilde{q}$. We will now discuss how this mode behaves as we increase $\tilde{q}$. Recall that, for small $\tilde{q}$, our perturbative analysis showed that this mode is unstable (exponentially growing) in the region of the RNdS parameter space shown in Fig.~\ref{fig:moduli}.

\begin{figure}[ht]
	\centering
	\includegraphics[width=\textwidth]{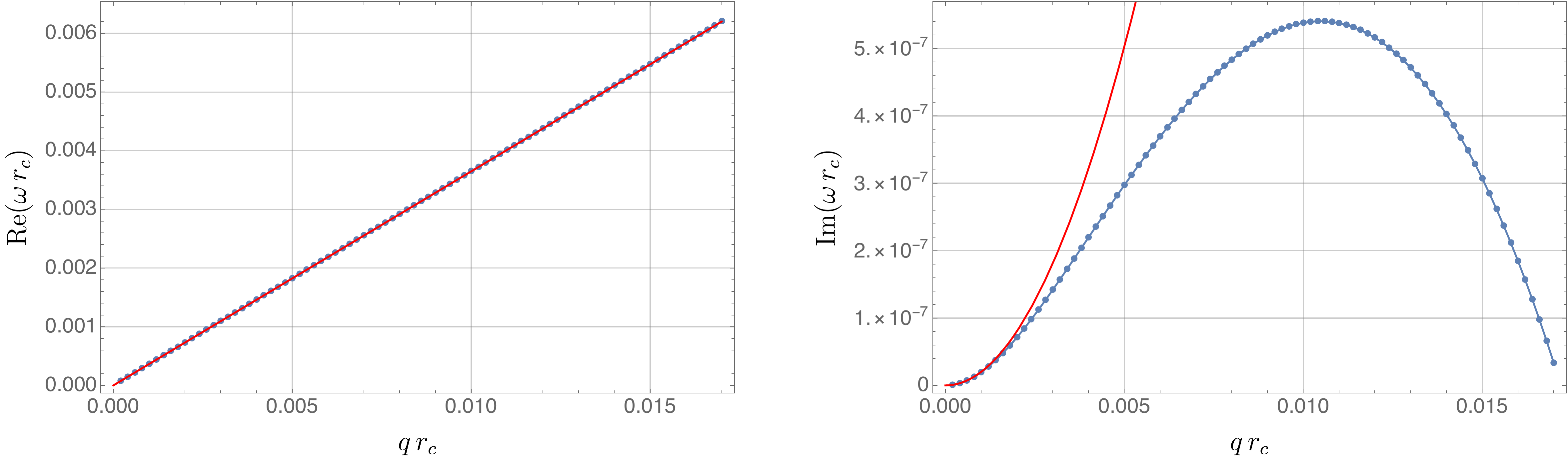}
	\caption{Unstable mode with $\ell=0$ as a function of $\tilde{q}=qr_c$ for $y_+=1/3$ and $Q/Q_{\mathrm{ext}}=1-10^{-3}$. The left panel shows the real part of $\mathrm{Re}(\omega\,r_c)$ and the right panel $\mathrm{Im}(\omega\,r_c)$. The dashed red line is the perturbative prediction of \eqref{omega1} (left panel) and (\ref{eq:secondorder}) (right panel) valid for small $\tilde{q}$. The blue dots are our numerical data. }
	\label{fig:small}
\end{figure}

We find that the unstable region of the RNdS moduli space presented in Fig.~\ref{fig:moduli} gets smaller as $\tilde{q}$ increases. In particular, we cannot find an instability beyond the region depicted in Fig.~\ref{fig:moduli}. We also find that the perturbative expansion in $\tilde{q}$ developed in section \ref{sec:smallq} works rather well at small $\tilde{q}$. For instance, in Fig.~\ref{fig:small} we show data for $y_+=1/3$ and $Q/Q_{\mathrm{ext}}=999/1000$ and $\ell=0$. The dashed red line indicates the analytic prediction of equations \eqref{omega1} and (\ref{eq:secondorder}), and the blue dots are the numerical data. The agreement at small $\tilde{q}$ is very reassuring.\footnote{As a further check on our numerical results, we have reproduced the results reported in Table I of Ref. \cite{Konoplya:2014lha}.} 

\begin{figure}[ht]
	\centering
	\includegraphics[width=0.63\textwidth]{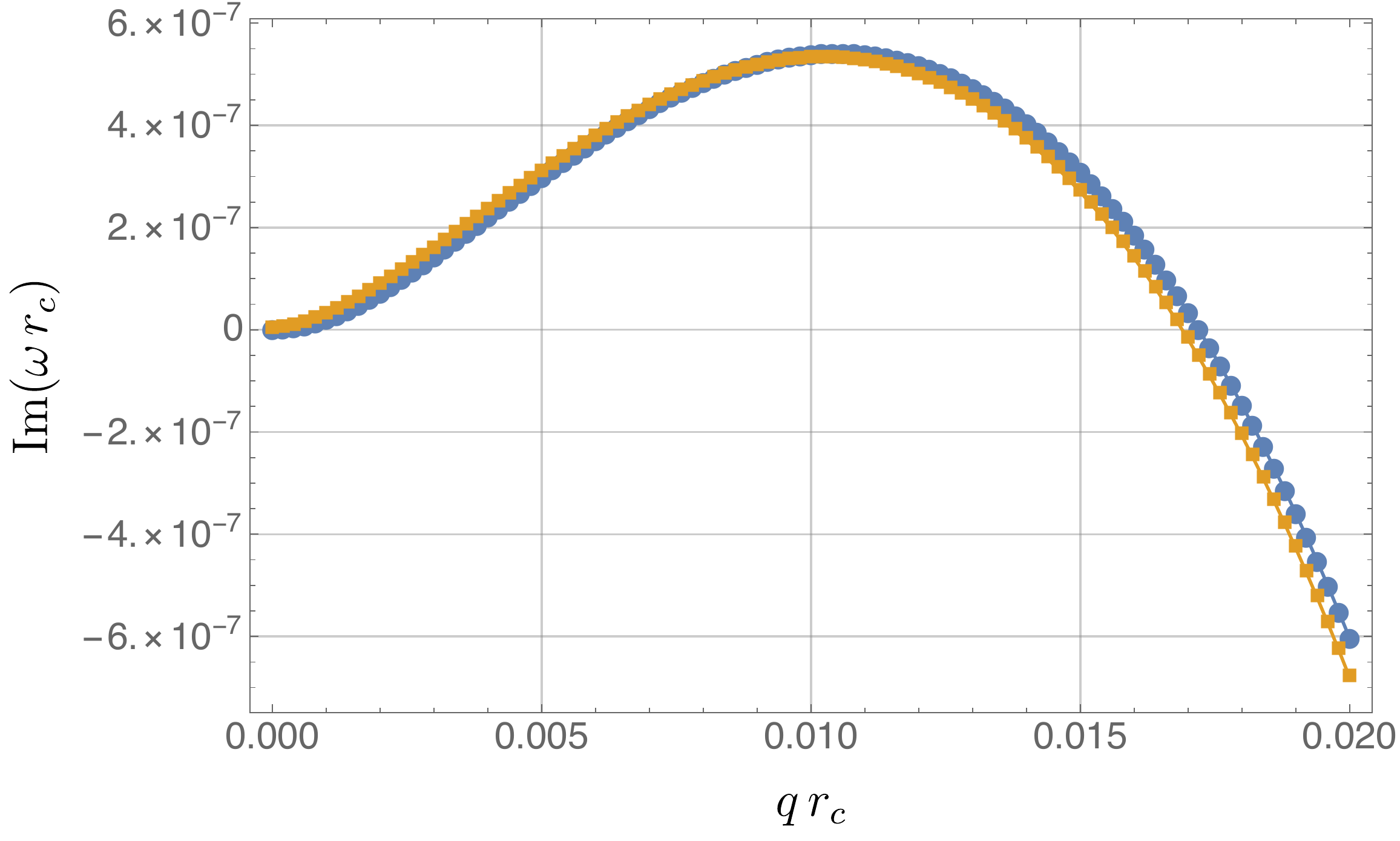}
	\caption{Unstable mode with $\ell=0$ for $y_+=1/3$ as a function of $\tilde{q}$ with $Q/Q_{\mathrm{ext}}=1-10^{-3}$ corresponding to the blue dots and $Q/Q_{\mathrm{ext}}=1-10^{-4}$ to the orange squares.}
	\label{fig:small2}
\end{figure}

We see from Fig. \ref{fig:small} that the instability shuts off as we increase $\tilde{q}$. The closer the black hole is to extremality the smaller the range of unstable $\tilde{q}$, although the unstable range of $\tilde{q}$ seems to remain non-zero at extremality. For instance, in Fig.~\ref{fig:small2} we plot the imaginary part of the unstable mode as a function of $\tilde{q}$ for $y_+=1/3$ and $Q/Q_{\mathrm{ext}}=1-10^{-3}$ and $Q/Q_{\mathrm{ext}}=1-10^{-3}$ (the blue dots) and $Q/Q_{\mathrm{ext}}=1-10^{-4}$ (orange squares).

\begin{figure}[bh]
	\centering
	\includegraphics[width=\textwidth]{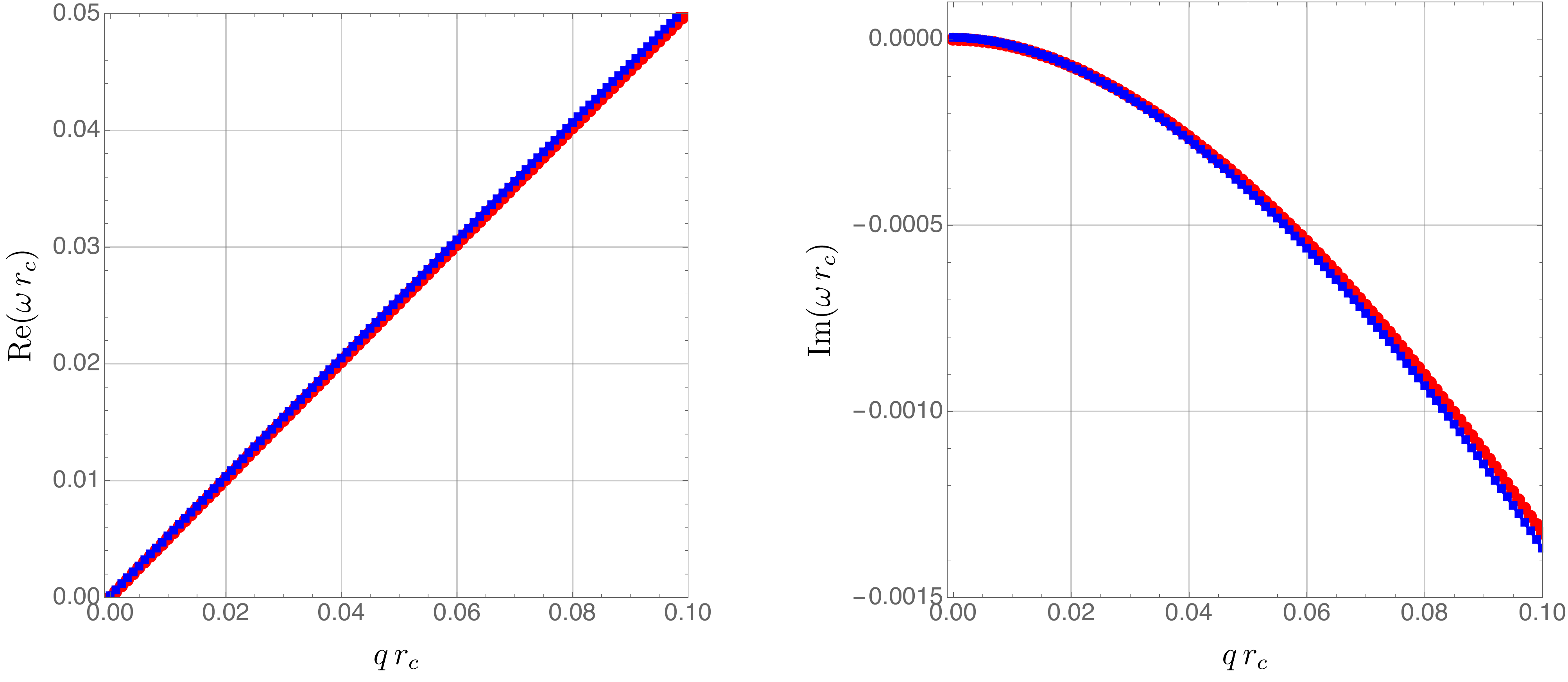}
	\caption{Quasinormal mode with $\ell=0$ for $y_+=1/2$ as a function of $\tilde{q}$ with $Q/Q_{\mathrm{ext}}=1-10^{-2}$ corresponding to the red dots and $Q/Q_{\mathrm{ext}}=1-10^{-3}$ to the blue squares.}
	\label{fig:agg2}
\end{figure}

For $y_+>\sqrt{2}-1$ we find no instability, no matter how close to extremality we get (we probed all the way down to $\sigma =10^{-7}$ with $\sigma$ defined in \eqref{NHregime2}). This is in excellent agreement with the small $\tilde{q}$ analysis of section \ref{sec:smallq}. For instance, in Fig.~\ref{fig:agg2} we plot the real (left panel) and imaginary (right panel) of the $\ell=0$ mode connecting to the trivial mode at small $\tilde{q}$, as a function of $\tilde{q}$. This data was collected for $y_+=1/2$ and $Q/Q_{\mathrm{ext}}=1-10^{-2}$ (the red dots) and $Q/Q_{\mathrm{ext}}=1-10^{-3}$ (the blue squares). Most importantly, the family of modes connected to the trivial mode when $\tilde{q}=0$, does not give $\mathrm{Im}(\omega)\to0$ as we approach extremality. This means that for any fixed $\tilde{q}$, when sufficiently close to extremality and when $y_+>\sqrt{2}-1$, the near-extremal mode discussed in section \ref{sec:NHqnmcharged} will always be dominant, i.e., more slowly decaying. 

\begin{figure}[ht]
	\centering
	\includegraphics[width=\textwidth]{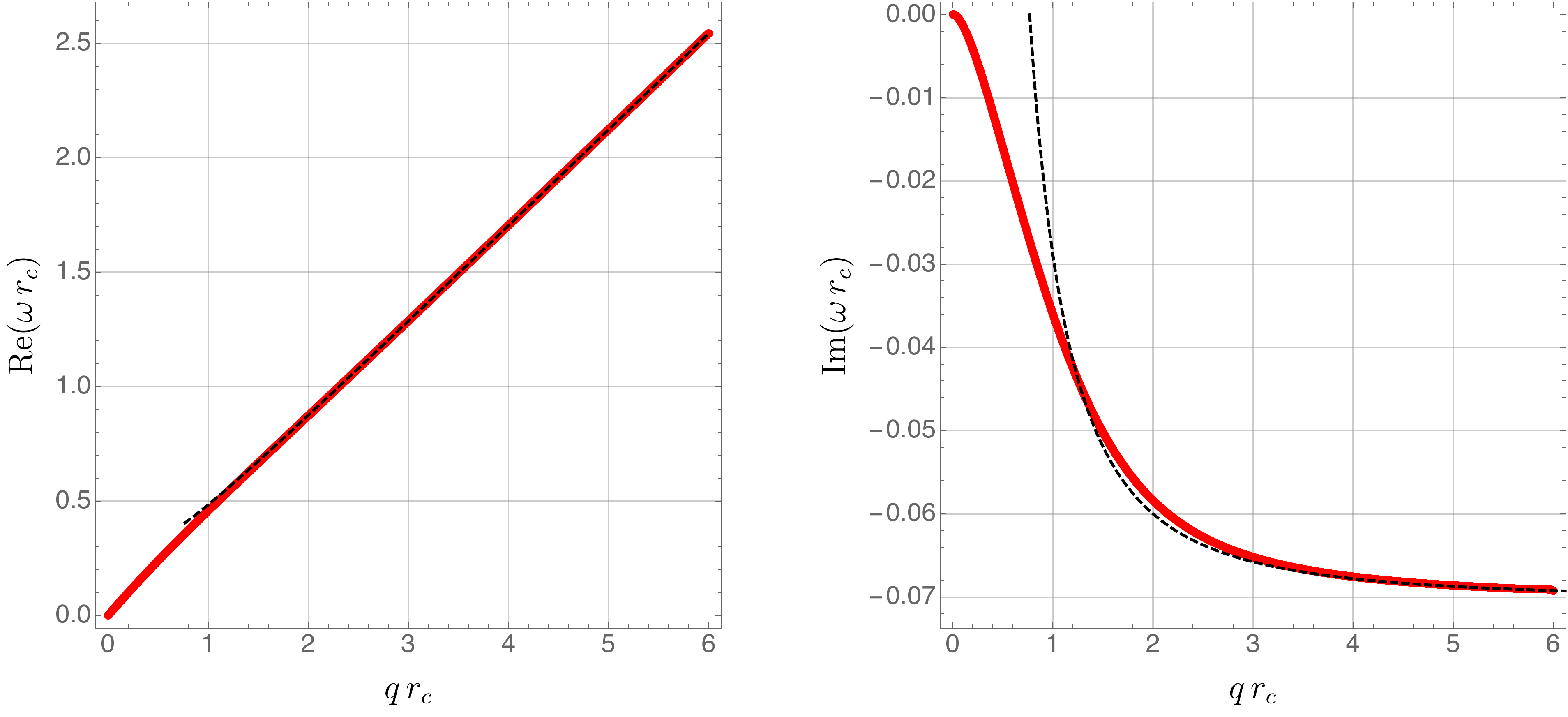}
	\caption{Quasinormal mode with $\ell=\mu=0$ as a function of $\tilde{q}$ for $y_+=1/2$ and $Q/Q_{\mathrm{ext}}=1-10^{-2}$, as a function of $\tilde{q}$. The red curve is our numerical data and the black line is the WKB prediction of equation \eqref{eqs:wkbsc}.}
	\label{fig:wkbmatch}
\end{figure}

At large $\tilde{q}$, this mode connects to the ``cosmological" WKB mode described by Eqs.~(\ref{eqs:wkbsc}). This is shown in Fig.~\ref{fig:wkbmatch}, which was collected with $y_+=1/2$ and $Q/Q_{\mathrm{ext}}=1-10^{-2}$. This is always the case, regardless of the magnitude of $y_+$, that is to say, regardless of whether  the mode being stable or unstable at small $\tilde{q}$.

\subsection{Photon sphere and de Sitter modes}

Ref. \cite{Cardoso:2017soq} classified uncharged scalar quasinormal modes into three familes: ``photon sphere", ``near-extremal" and "de Sitter".\footnote{The special mode discussed in the previous section reduces to the trivial mode when $q=0$, which lies outside this classification.} We will see in the next section that near-extremal modes are the ones most relevant for strong cosmic censorship. Here we will discuss briefly what happens to the photon sphere and de Sitter modes as one increases the charge, starting from $q=0$.

\begin{figure}[ht]
	\centering
	\includegraphics[width=\textwidth]{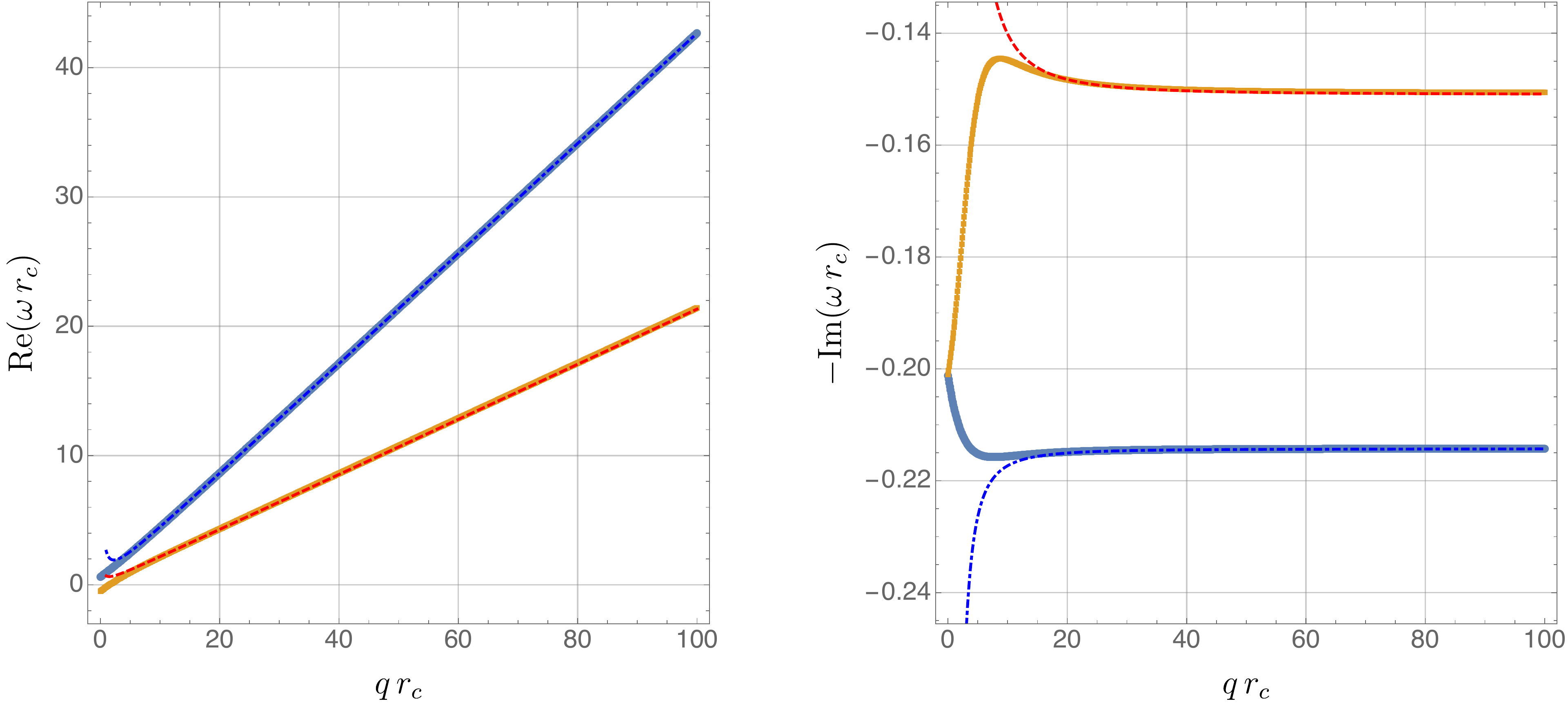}
	\caption{Photon sphere quasinormal mode with $\ell=\mu=0$ as a function of $\tilde{q}$ for $y_+=1/2$ and $Q/Q_{\mathrm{ext}}=1/2$, as a function of $\tilde{q}$. The dots are our numerical data, the dash-dotted blue line it the ``black hole'' WKB prediction Eqs.~\eqref{eqs:wkbs} and the dashed red line the ``cosmological'' WKB prediction Eqs.~\eqref{eqs:wkbsc}.}
	\label{fig:split}
\end{figure}

First consider the photon sphere modes of \cite{Cardoso:2017soq}. When $\tilde{q}=0$, we have two modes with equal imaginary part, and real parts of equal magnitude but opposite sign. This is just a consequence of complex conjugation (as discussed at the end of section \ref{sec:chargedScalarSCC}), which is a symmetry when $\tilde{q}=0$. However, this degeneracy is broken when we consider $\tilde{q}\neq0$. In Fig.~\ref{fig:split} we plot data, as a function of $\tilde{q}$, for $y_+=1/2$ and $Q/Q_{\mathrm{ext}}=1/2$. At small $\tilde{q}$ we find that one photon sphere mode is a mode that approaches, in the large $\tilde{q}$ regime, the ``cosmological'' WKB prediction (represented as the dashed red line), and the other photon sphere mode approaches the ``black hole'' WKB prediction (represented as the dash-dotted blue line).

\begin{figure}[ht]
	\centering
	\includegraphics[width=\textwidth]{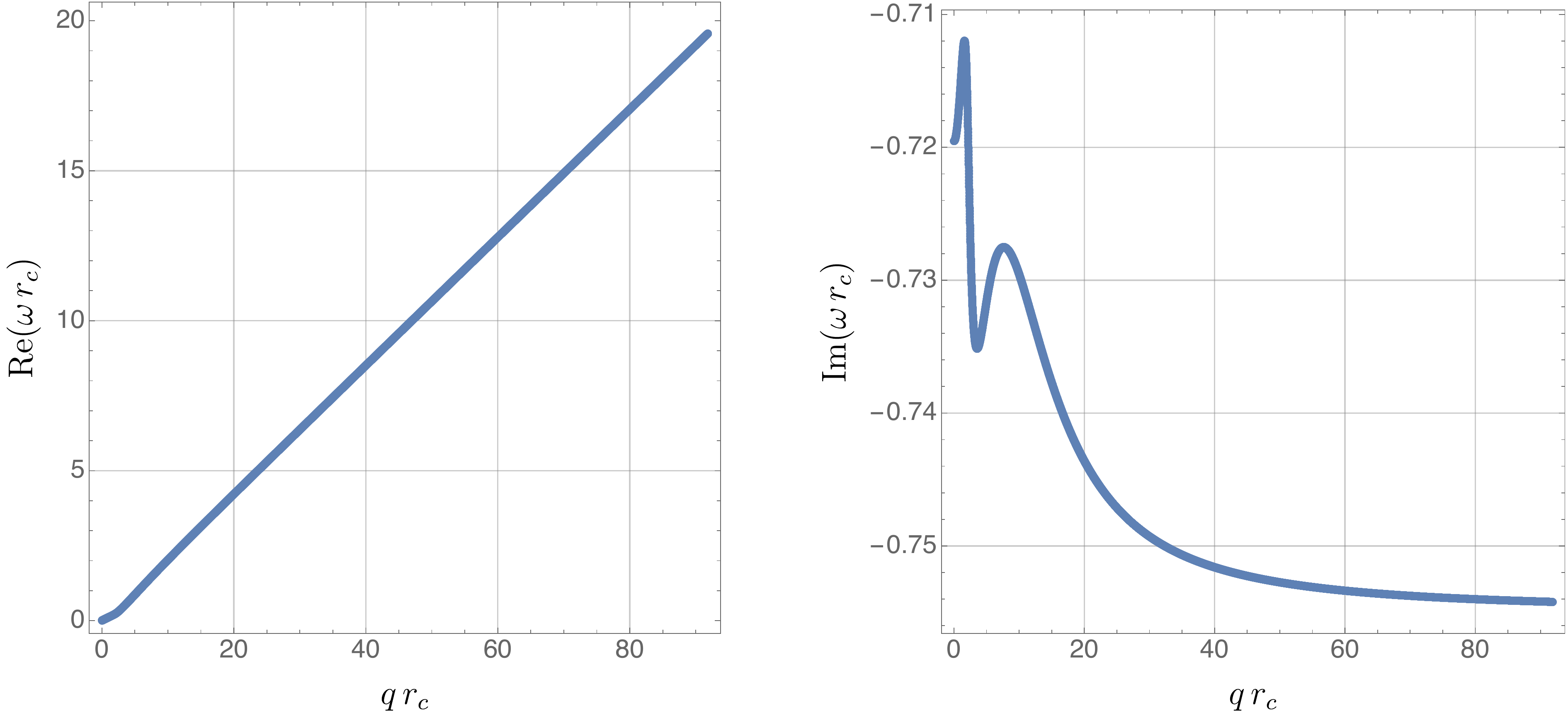}
	\caption{De Sitter quasinormal mode with $\ell=1$ and $\mu=0$ as a function of $\tilde{q}$ for $y_+=1/2$ and $Q/Q_{\mathrm{ext}}=1/2$, as a function of $q\,r_c$.}
	\label{fig:wkbnasty}
\end{figure}

Finally, we have also studied the fate of the de Sitter modes of \cite{Cardoso:2017soq}. At $\tilde{q}=0$ this mode is purely imaginary, but acquires a real part when $\tilde{q}\neq0$. At large $\tilde{q}$ the mode seems to approach a constant which does not seem to be captured by our WKB analysis. This is exemplified in Fig.~\ref{fig:wkbnasty} which was determined for $y_+=1/10$, $Q/Q_{\mathrm{ext}}=1/2$, $\ell=1$ and $\tilde{\mu}=0$. Most importantly, this constant remains non-zero as we approach extremality, which implies that these modes are not relevant for strong cosmic censorship.

%%%%%%%%%%%%%%%%%%%%%%%%%%%%%%%%
\subsection{\label{sec:resultscsm}Near extremal modes}
%%%%%%%%%%%%%%%%%%%%%%%%%%%%%%%%
In this section we will discuss our numerical results for large $\tilde{q}\equiv qr_c $, focusing on the near-extremal family of quasinormal modes discussed in section \ref{sec:NHqnmcharged} since these are the slowest decaying modes and therefore the most relevant for strong cosmic censorship. The main goal of this section is to show that, no matter how large $\tilde{q}$ is, we always find regions of the RNdS parameter space where a violation of strong cosmic censorship occurs. We will show this in two ways: first, consider a fixed near-extremal RNdS black hole and increase $\tilde{q}$. Second we will take $\tilde{q}$ fixed, and let the black hole approach extremality. In both cases we will see the same effect emerging. 

We remind the reader that a gauge transformations of the form $\tilde{A} = A-\check{\phi}\,\mathrm{d}t$ (with constant $\check{\phi}$) transforms $\Phi\sim e^{-i\,\omega\,t}$ to $\tilde{\Phi} \sim e^{-i\,\tilde{\omega}\,t}$, where $\tilde{\omega}=\omega+q\,\check{\phi}$. For this reason, it is convenient to plot 
\be
\varpi\equiv \omega-q\,A_{t}(r_+)=\omega-q\, Q/r_+
\label{varpidef}
\ee
in what follows, since this quantity is invariant under such large gauge transformations.

The WKB expansion of section \ref{sec:WKBqnmcharged} is valid in the entire range of RNdS parameters. As discussed above, the WKB result demonstrates that, for any fixed RNdS black hole, one can achieve $\beta<1/2$ by taking sufficiently large $\tilde{q}$. However, for fixed $\tilde{q}$, we explained that non-perturbative effects might become important for near-extremal black holes. We will therefore focus our numerical efforts on near-extremal RNdS black holes. In order to identity the lowest lying quasinormal mode, we have run extensive eigenvalue searches and we always found that the near extremal mode discussed above dominates when $\sigma\ll1$ ($\sigma$ is defined in \eqref{NHregime2}). Once the mode was identified, a standard Newton-Raphson algorithm was employed.

\begin{figure}[ht]
	\centering
	\includegraphics[width=0.98\textwidth]{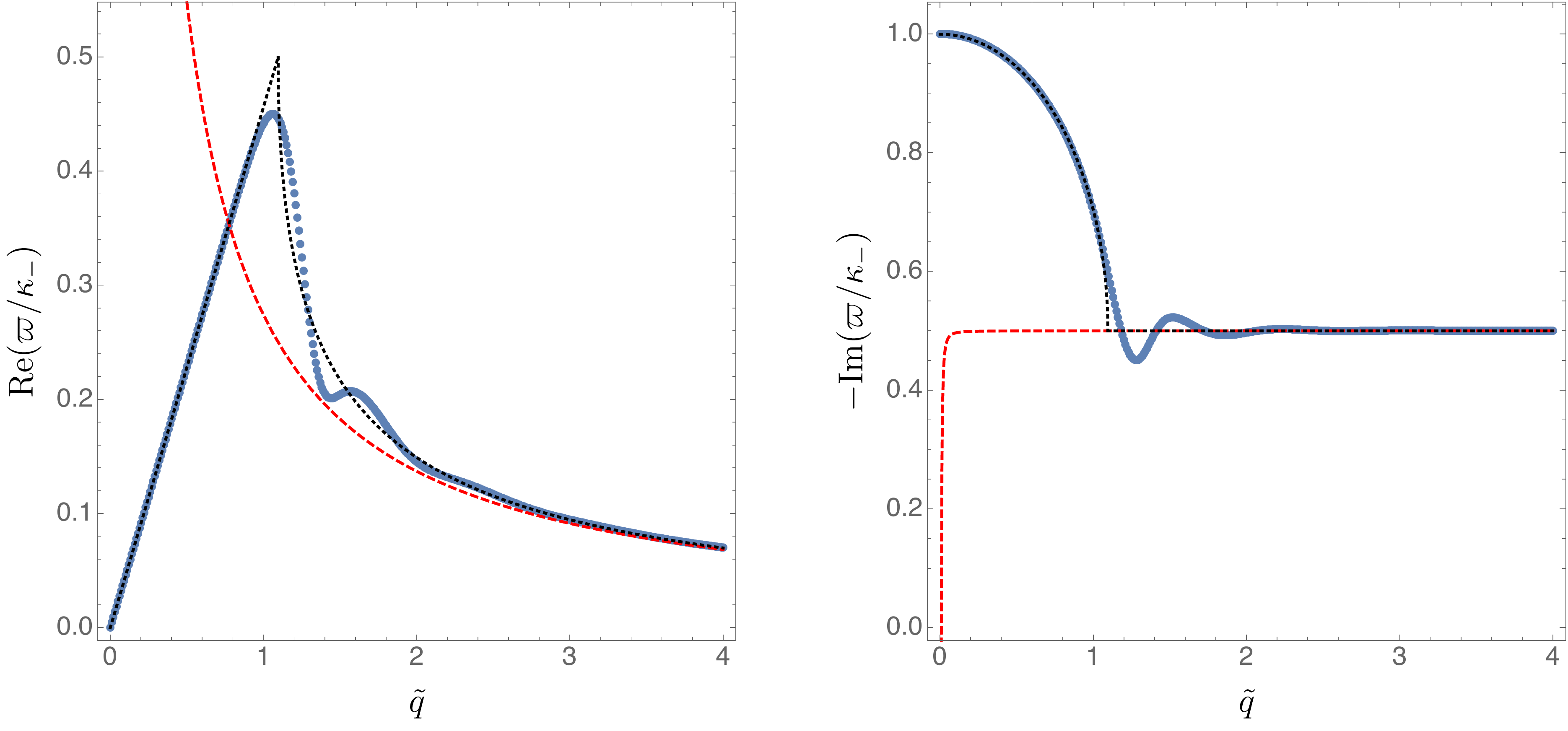}
	\caption{Real part (left panel) and imaginary part (right panel) of the near-extremal quasinormal mode with zero overtone as a function of $\tilde{q}$. This data was generated with $Q/Q_{\mathrm{ext}}=1-10^{-4}$, $y_+=1/3$, $\mu=0$ and $\ell=0$. The blue dots are our numerical data, the dashed red line shows the WKB prediction \eqref{eqs:wkbs} and the black dotted line is the near horizon prediction \eqref{eq:nhsum}.} 
	\label{fig:fixedQ}
\end{figure}

In Fig.~\ref{fig:fixedQ} we plot both the real part (left panel) and imaginary part (right panel) of the near-extremal quasinormal mode with zero overtone, and hold fixed $Q/Q_{\mathrm{ext}}=1-10^{-4}$, $y_+=1/3$, $\tilde{\mu}=0$ and $\ell=0$. We see that $-{\rm Im}(\omega)/\kappa_-$ initially decreases rather rapidly with $\tilde{q}$ until we reach $\tilde{q}\simeq \check{q}_c$. Sufficiently close to extremality we have $\check{q}_c=\tilde{q}_c$ (with $\tilde{q}_c$ given as in \eqref{eq:qc}). The dashed red line in Fig.~\ref{fig:fixedQ} indicates the WKB prediction of \eqref{eqs:wkbs} and the dotted black line the near horizon prediction \eqref{eq:nhsum}. Note that the large $\tilde{q}$ behaviour of the near horizon expansion agrees with the near extremal WKB result in the same limit.

\begin{figure}[ht]
	\centering
	\includegraphics[width=1\textwidth]{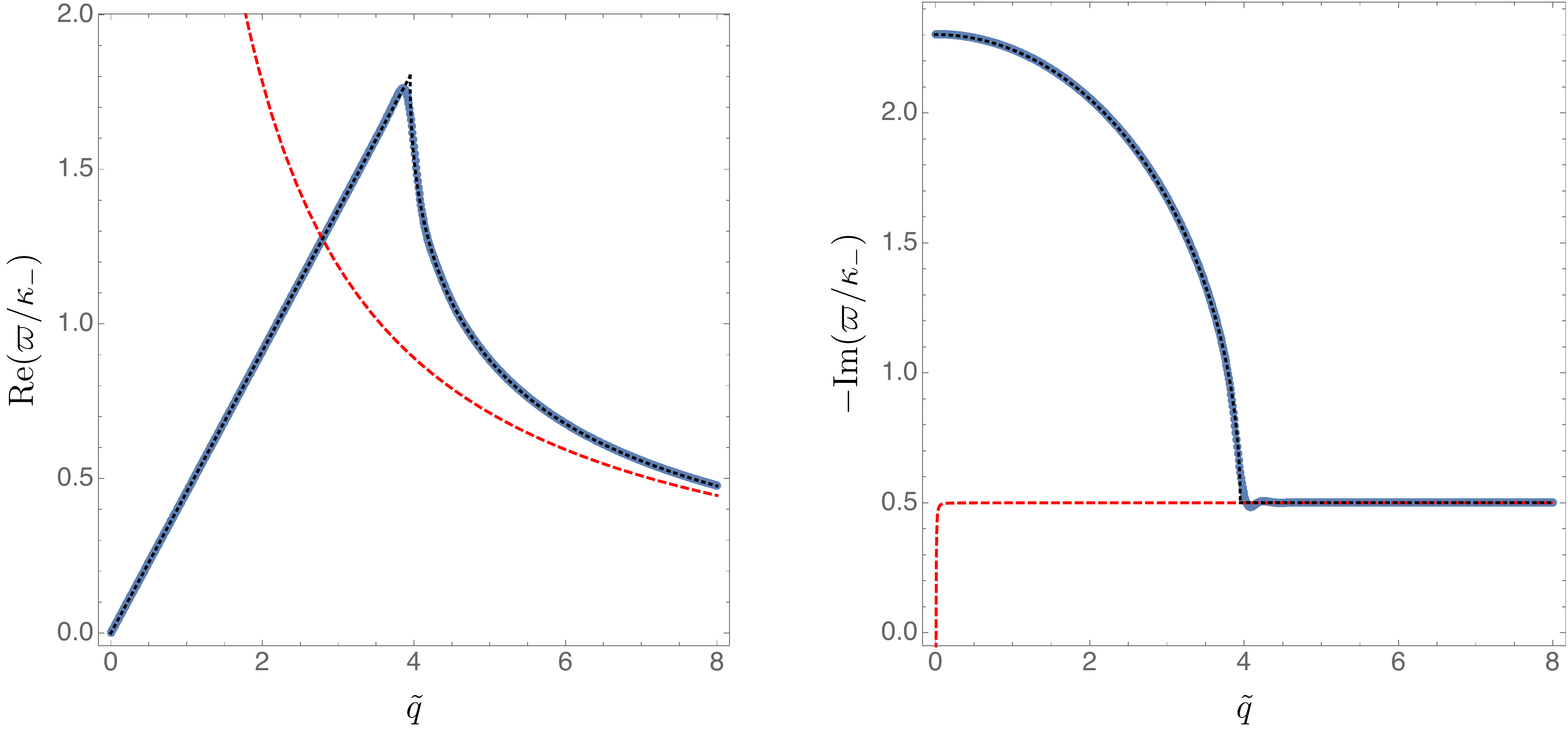}
	\caption{The same as Fig. \ref{fig:fixedQ} except that $\ell=1$.}
	\label{fig:fixedQell}
\end{figure}

Similar results hold if we change the black hole parameters, or the perturbation parameters. For instance, in Fig.~\ref{fig:fixedQell} we keep the same parameters as those used to generate Fig.~\ref{fig:fixedQ}, but instead take $\ell=1$.

\begin{figure}[ht]
	\centering
	\includegraphics[width=0.55\textwidth]{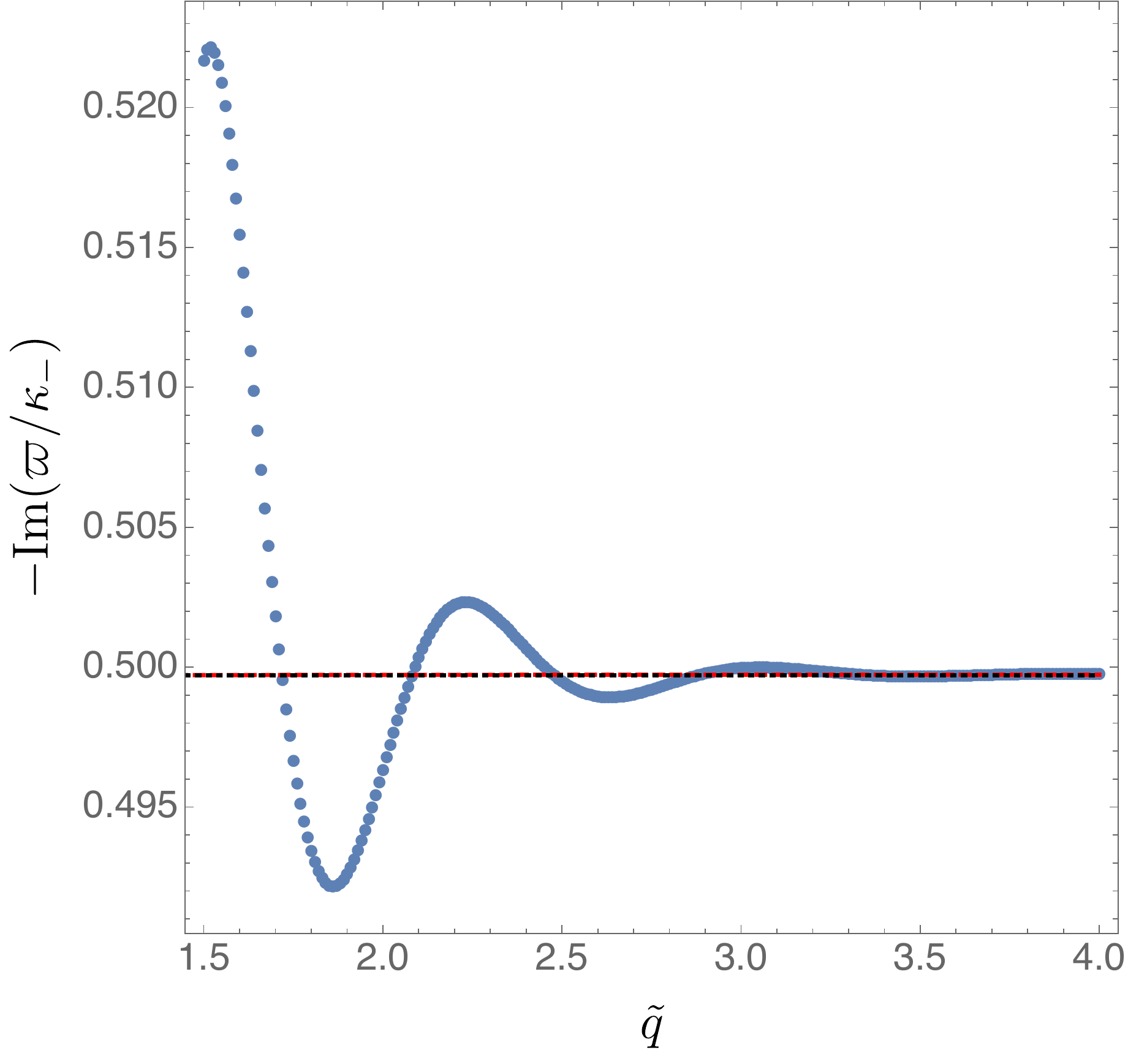}
	\caption{Zoom of the right panel of  Fig.~\ref{fig:fixedQ} near the leading WKB prediction. The red dashed line is the leading WKB prediction, and the blue dots are our numerical data.}
	\label{fig:wiggles}
\end{figure}

An eagle-eyed reader would have not failed to notice that the large $\tilde{q}$ behaviour of our numerics does not seem to match exactly the WKB prediction outlined in section \ref{sec:WKBqnmcharged}. In Fig.~\ref{fig:wiggles} we zoom near the WKB prediction, and observe that there are ``wiggles", which decrease in size with increasing $\tilde{q}$, and that are centred around the leading WKB prediction. These wiggles can be shown to decrease exponentially fast in $\tilde{q}$ and are precisely the type of non-perturbative effect that an asymptotic series, such as WKB, cannot easily capture. It is tempting to fit the difference between the data for $-{\rm Im}(\omega)/\kappa_-$ and the WKB approximation with an \emph{ansatz} of the form
\be
\Delta = a_0\,e^{-b_0\,q^\alpha}\sin(\Omega_0\,q^{\theta}+\beta)\,.
\label{eq:ansatz}
\ee
Such an \emph{ansatz} does seem to fit well the wiggles we observe. However, we find that the values we extract for the fit parameters are sensitive to where we start the fit.

\begin{figure}[ht]
	\centering
	\includegraphics[width=0.45\textwidth]{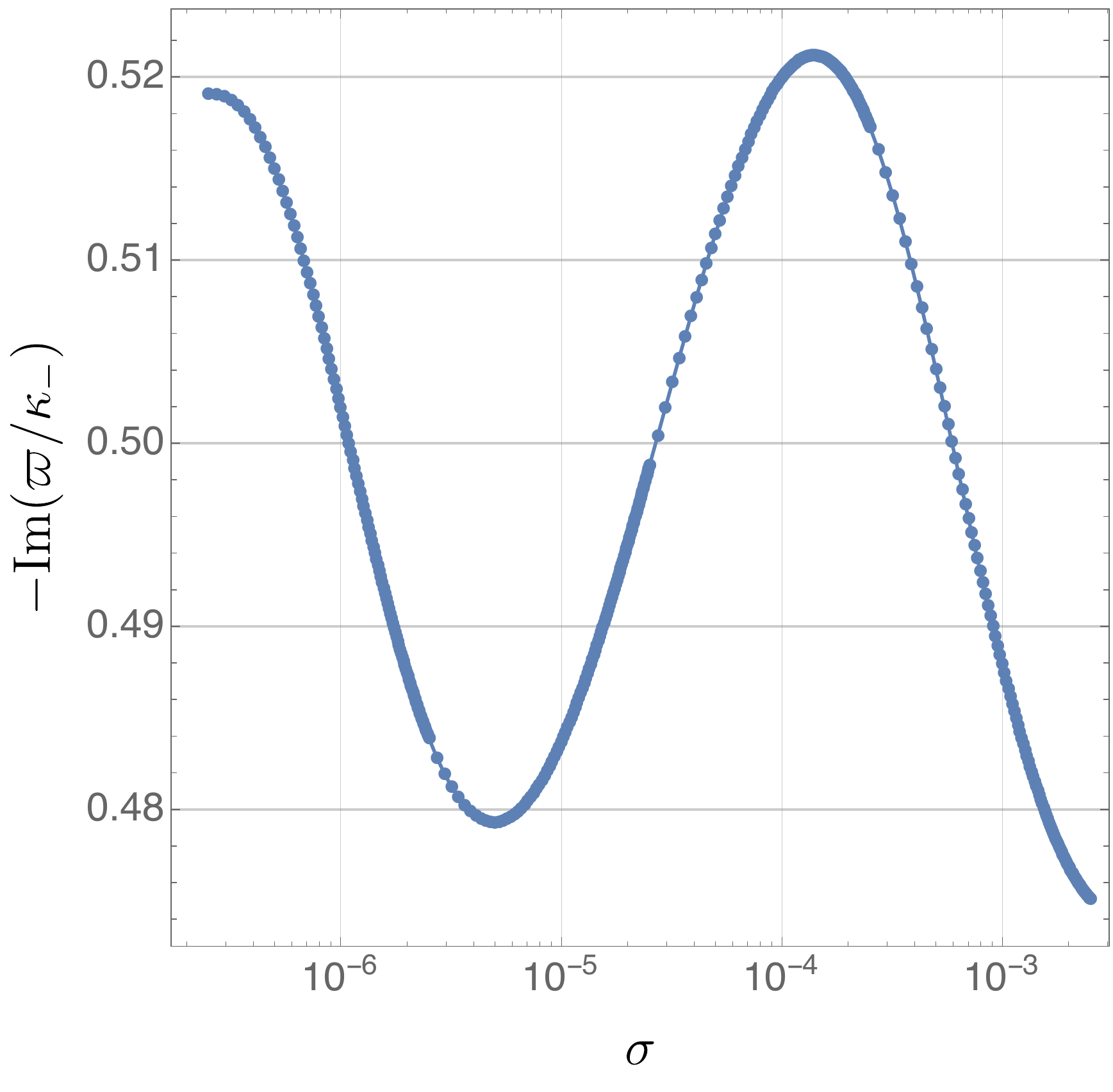}
	\caption{A linear-log plot of $-{\rm Im}(\varpi)/\kappa_-$ as a function of $\sigma \equiv 1-r_-/r_+$ for $\ell = 0$, $y_+=1/2$, $\tilde{q}=0.75$ and $\mu=0$.}
	\label{fig:wiggles2}
\end{figure}

The fact that the phase in \eqref{eq:ansatz} is not universal, shows that these wiggles should appear as we move in other directions of parameter space. In particular, if we keep $\tilde{q}$ fixed and move towards extremality in the RNdS moduli space then we expect the wiggles to be present. In Fig.~\ref{fig:wiggles2} we plot $-{\rm Im}(\omega)/\kappa_-$ as a function of the non-extremality parameter $\sigma$ and show that the approach to extremality in not monotonic for $\tilde{q}>\tilde{q}_c$. In this figure we use $\ell = 0$, $y_+=1/2$, $\tilde{q}=0.75$ (for these parameters $\tilde{q}_c\approx 0.533002$) and $\tilde{\mu}=0$.

For $\sigma<10^{-5}$, these data seem to be accurately described by a discrete-self similar model of the form
\be
-\frac{{\rm Im}(\omega)}{\kappa_-} = a_0 +b_0\,\cos(\Omega_0\log \sigma+c_0)\,,
\ee
with the fit parameters $a_0, b_0, c_0$ depending on $y_+$ and the scalar field parameters. For instance, for the parameters used in generating Fig.~\ref{fig:wiggles2} we find
\be
-\frac{{\rm Im}(\omega)}{\kappa_-} \approx 0.49915-0.020340 \cos \left(1.0680 \log \sigma +0.48503\right)\,.
\ee
One might wonder why these wiggles were missed by the near horizon analysis of section \ref{sec:NHqnmcharged}. We note that the wiggles arise for parameters where the matching conditions of section \ref{sec:NHqnmcharged} were very poorly motivated. To recover the wiggles, one would need to determine an outer solution and then match to the inner solution we found. This kind of matching has been accomplished in \cite{Yang:2013uba} in the context of gravitational perturbations around asymptotically flat Kerr black holes. Near extremality, wiggles were observed that seem analogous to what we are finding.

What are the consequences of the wiggles for strong cosmic censorship? Our results indicate that, even for large scalar field charge $\tilde{q}>\tilde{q}_c$ there are near-extremal black holes for which $\beta>1/2$ and so the Christodoulou version of strong cosmic censorship is violated. However, unlike the uncharged case, there is no neighbourhood of extremality in which {\it all} black holes violate strong cosmic censorship. Furthermore, for large $\tilde{q}$, it is clear that the wiggles have very small amplitude. This has two related consequences. First, as one approaches extremality at fixed large $\tilde{q}$, the value of $\beta$ can exceed $1/2$ but only by a tiny amount. Second, the larger $\tilde{q}$ is, the closer one has to get to extremality in order for the wiggles to overcome the leading WKB prediction, i.e., the closer one has to get to extremality in order to violate strong cosmic censorship. For physical values of the scalar field and black hole parameters we have $\tilde{q} \ggg \tilde{q}_c$ (see below \eqref{eq:qc}) and so one would have to get incredibly close to extremality to see a violation of strong cosmic censorship.\footnote{If one is interested in strong cosmic censorship only as a mathematical statement then one can see that it is violated for any given $q$ by taking $r_+$ small enough to achieve $q_c>q$, in which case $\beta$ becomes significantly larger than $1/2$. For a physical value of the scalar field charge this would correspond to a sub-Planckian black hole.} In these senses, the violation of strong cosmic censorship exhibited in the presence of a charged scalar field is milder than the violation present with an uncharged scalar field and certainly nowhere near as bad as what happens for gravitoelectromagnetic perturbations. 

Our numerical results above are for $\mu=0$. Turning on $\mu$ does not change much the overall picture, except that for sufficiently large $\mu$ the instability disappears from the spectrum, and in order for the WKB approximation to work one needs to move towards larger values of $\tilde{q}$. For fixed $q<q_c$, increasing $\mu$ will tend to increase $\beta$. In addition, as long as $\mu\neq 0$, the near horizon expansion \eqref{eq:deltaomegaminus} for $q=0$ predicts that if we are at extremality and send $y_+\to 1$ then $\beta \to\infty$ as 
$$
\beta = \sqrt{\frac{3}{2}} \frac{\tilde{\mu }}{\sqrt{1-y_+}}.
$$
This divergence of $\beta$ at extremality when $y_+\to 1$ is similar to happens in the case of gravitoelectromagnetic perturbations \cite{Dias:2018etb}. 
%In addition, our near horizon analysis predicts that, for fixed $q<q_c$, increasing $\mu$ will tend to increase $\beta$.

One might wonder why wiggles were not observed for Kerr-dS in Ref. \cite{Dias:2018ynt}. The charge $q$ is analogous to the azimuthal quantum number $m$ of perturbations of Kerr-dS. Based on this analogy, we believe that the wiggles should also be present for Kerr-dS. Since $m$ is an integer, it cannot be taken small enough to make the wiggles apparent in the numerical data, which is why they were not discovered in Ref. \cite{Dias:2018ynt}. Even if these wiggles are present for Kerr-dS, it does not invalidate the conclusion of Ref. \cite{Dias:2018ynt}. This is because, for Kerr-dS, $m$ can be arbitrarily large. The larger $m$ is, the closer to extremality one has to go in order to see the wiggles. In particular, as $m\to+\infty$ there should be no wiggles at all. In other words, the WKB prediction should work for Kerr-dS because $m$ can be taken arbitrarily large. This is in constrast with RNdS, for which the scalar field charge $q$ is a fixed parameter. 

The analogy with Kerr-dS does suggest a way of recovering strong cosmic censorship (with smooth initial data) for RNdS. Instead of a single scalar field of charge $q$ we could consider an infinite tower of scalar fields with charges $q_n$ and masses $\mu_n$. This happens for example in Kaluza-Klein theory. If $q_n \rightarrow \infty$ as $n \rightarrow \infty$ then we might expect the WKB approximation to become exact, in which case strong cosmic censorship would be enforced. 

%%%%%%%%%%%%%%%%%%%%%%%%%%%%%%%%
\subsection{Comparison with other work}
%%%%%%%%%%%%%%%%%%%%%%%%%%%%%%%%
Our charged scalar results seem to be in tension with those of Ref. \cite{Hod:2018dpx} where it was claimed that for sufficiently large scalar field mass and charge, strong cosmic censorship would be recovered. The analysis of Ref. \cite{Hod:2018dpx} assumes
\be
\label{hod_assume}
\mu\,r_+\ll q\,Q\ll\mu^2 r_+^2\,.
\ee
Note that for near-extremal black holes $Q \sim r_+$ so these assumptions imply $q \gg \mu$ and $r_+ \gg q/\mu^2$. Therefore the analysis of Ref. \cite{Hod:2018dpx} applies to near-extremal holes only when they are sufficiently large.\footnote{Supermassive, if we take $q$ and $\mu$ to be equal to the charge and mass of the electron.} For parameters satisfying (\ref{hod_assume}), Ref. \cite{Hod:2018dpx} predicts that the slowest decaying quasinormal mode has frequency
\be
\label{hod_predict}
 \omega = \frac{qQ}{r_+} + \frac{\mu^2 r_+^2 \kappa_+}{2qQ}- i \frac{\kappa_+}{2} \left( 1- \frac{qQ}{\mu^2 r_+^2} \right).
\ee
The conditions (\ref{hod_assume}) are not satisfied by our data above (since we have set $\mu=0$ in Fig.~\ref{fig:wiggles2}). The regime (\ref{hod_assume}) is a particularly difficult corner of parameter space to study numerically, since it requires two distinct hierarchy of scales. We have computed the zero overtone near-extremal quasinormal mode with $\ell=0$, $y_+=1/2$, $\tilde{Q}=2/5$, $q\,Q=1000$ and $\mu\,r_c=200$ (so $\mu \,r_+ = 100$). For these parameters we find numerically that ($\tilde{\varpi} = \varpi \, r_c$ with $\varpi$ defined in (\ref{varpidef})) 
\be
\tilde{\varpi}=0.34316055 - 0.034273947\,i\,.
%\tilde{\varpi}=0.3431605519292390 - 0.0342739466983198\,i\,.
\ee
Our ``black hole'' WKB analysis (see Eqs.~(\ref{eqs:wkbs})) seems to be in good agreement with the numerics: for this parameter choice it gives
\be
\tilde{\varpi}_{\mathrm{WKB}}=0.34286131 - 0.034273959 \,i\,.
%\tilde{\varpi}_{\mathrm{WKB}}=0.3428613061224490 - 0.0342739592626006 \,i\,.
\ee
On the other hand, the prediction of (\ref{hod_predict}) is not so good for the imaginary part:
\be
\tilde{\varpi}_{\rm Hod}=0.34285949 - 0.030857166\,i\,.
%\tilde{\varpi}_{\rm Hod}=0.3428594938775510 - 0.0308571663671857\,i\,.
\ee
However, if we discard the ${\cal O}(10 \%)$ correction to the imaginary part in (\ref{hod_predict}) then we obtain much better agreement:
\be
\tilde{\varpi}_{\rm Hod, leading}=0.34285949 - 0.034285714\,i\,.
%\tilde{\varpi}_{\rm Hod, leading}=0.3428594938775510 - 0.03428571428571\,i\,.
\ee
This suggests that there may be a mistake in Ref. \cite{Hod:2018dpx} and that the subleading correction to the imaginary part is much smaller than stated in \eqref{hod_predict}. This would imply that the analysis of Ref. \cite{Hod:2018dpx} is inconclusive for strong cosmic censorship, even in the region of parameter space where this calculation is valid. In any case, the analysis of \cite{Hod:2018dpx} is a WKB analysis which will miss non-perturbative effects. As we have explained above, it is non-perturbative effects that are responsible for ensuring that the violation of strong cosmic censorship first observed in \cite{Cardoso:2017soq} persists when the scalar field has a non-zero charge.

We can also compare our results with those of Refs. \cite{Cardoso:2018nvb,Mo:2018nnu}, which appeared when this work was almost finished. Our results appear to be in agreement with the numerical results of these papers where there is overlap. 
However, we disagree with their conclusion that strong cosmic censorship is respected for large enough $q$. We have seen that the ``wiggles" discussed above always lead to a violation of strong cosmic censorship close enough to extremality. We believe that these papers did not discover the wiggles because they did not consider sufficiently near-extremal black holes. 

%%%%%%%%%%%%%%%%%%%%
\section{Discussion} 
\label{sec:discussion} 
%%%%%%%%%%%%%%%%%%%%

We have seen the the introduction of a charged scalar field does not rescue strong cosmic censorship for RNdS black holes. 
Even when the charge of the field is large, there is always a tiny neighbourhood of extremality in which strong cosmic censorship is violated. To rescue strong cosmic censorship it appears that one would have to allow rough initial data, as proposed in Ref. \cite{Dafermos:2018tha}.

Strong cosmic censorship is violated for large scalar field charge because of the ``wiggles" discussed above. It would be interesting to obtain an analytic understanding of this effect by performing a proper matching of the near horizon modes to an outer solution, as in Ref. \cite{Yang:2013uba}.

It seems unlikely that this violation of strong cosmic censorship has any significance for astrophysical black holes. To see a violation of strong cosmic censorship with a realistic value for the scalar field charge and a macroscopic black hole, the hole has to be incredibly close to extremality. However, highly charged black holes are not expected to occur in Nature. Furthermore, our analysis has been entirely classical. Quantum mechanically, a near-extremal RNdS black hole will evolve away from extremality via Hawking radiation of charged particles (see Ref. \cite{Dias:2018etb} for further discussion). 

%%%%%%%%%%%%%%%%%%%%%%%%%%%%%%%%%%%%%%%%%%%%%%%
\subsection*{Acknowledgments}

We are grateful to the authors of Ref. \cite{Mo:2018nnu} for explaining their results to us. 
OJCD is supported by the STFC Ernest Rutherford Grants No. ST/K005391/1 and No. ST/M004147/1, and by the STFC ``Particle Physics Grants Panel (PPGP) 2016" Grant No. ST/P000711/1. HSR and JES were supported in part by STFC Grants No. PHY-1504541 and ST/P000681/1.

\bibliographystyle{JHEP}
\bibliography{rnds}{}

\end{document}